\documentclass[preprint,12pt]{elsarticle}

\usepackage{graphicx}
\usepackage{dcolumn}
\usepackage{bm}
\usepackage{epstopdf, epsfig}
\usepackage{lmodern}
\usepackage{amsmath}
\usepackage{bm}
\usepackage{amssymb}
\usepackage{graphicx}

\usepackage{subcaption}
\usepackage{textcmds}
\usepackage{placeins}
\usepackage{dashrule}
\usepackage{color}
\usepackage{xcolor}
\usepackage{hyperref}
\hypersetup{
    colorlinks=true,
    linkcolor=blue,
    filecolor=blue,      
    urlcolor=black,
    citecolor = blue
    }
\usepackage{tikz}
\usepackage{blindtext}
\usepackage{times}
\usepackage{paralist}
\usepackage{MnSymbol} 
\usepackage{tikz}
\usepackage[export]{adjustbox}
\usepackage[most]{tcolorbox}
\usepackage{varwidth}   

\usepackage[normalem]{ulem}
\usepackage{color}

\newcommand{\rev}[1]{{#1}}

\newcommand{\alvaro}[1]{{\color{black}#1}}
\definecolor{gray}{rgb}{0.6,0.6,0.6}

\journal{International Journal of Heat and Fluid Flow}

\begin{document}
\begin{frontmatter}

\title{Data-driven assessment of arch vortices in simplified urban flows}

\author[label1,label3]{{\'A}lvaro Mart{\'\i}nez-S{\'a}nchez}
\author[label2]{Eneko Lazpita}
\author[label2]{Adri\'an Corrochano}
\author[label2]{Soledad Le Clainche}
\author[label1]{\rev{Sergio} Hoyas}
\author[label3]{Ricardo Vinuesa\corref{cor1}}

\cortext[cor1]{Corresponding author. e-mail: rvinuesa@mech.kth.se}

\address[label1]{Instituto Universitario de Matem\'atica Pura y Aplicada, 
Universitat Polit\`ecnica de Val\`encia, Valencia 46022, Spain.}
\address[label2]{School of Aerospace Engineering, Universidad Polit\'ecnica de Madrid, 28040 Madrid, Spain
}
\address[label3]{FLOW, Engineering Mechanics, KTH Royal Institute of Technology, SE-100 44 Stockholm, Sweden}

\begin{abstract}

Understanding flow structures in urban areas \rev{is widely} recognized as a challenging concern due to its effect on urban development, air quality, and pollutant dispersion. In this study, state-of-the-art data-driven methods for modal analysis of simplified urban flows are used to study the dominant flow processes \rev{in these environments}. Higher order dynamic mode decomposition (HODMD), a highly-efficient method to analyze turbulent flows, is used together with traditional techniques such as proper-orthogonal decomposition (POD) to analyze high-fidelity simulation data of a simplified urban environment. Furthermore, the spatio-temporal Koopman decomposition (STKD) will be applied to the temporal modes obtained with HODMD to perform spatial analysis. The flow interaction within the canopy influences the flow structures, particularly the arch vortex. The latter is a vortical structure generally found downstream of wall-mounted obstacles, which is generated as a consequence of flow separation. Therefore, the main objective of the present study is to characterize the mechanisms that promote these phenomena in urban areas with different geometries. Remarkably, among all the vortical structures identified by the HODMD algorithm, low- and high-frequency modes are classified according to their relation with the arch vortex. They are referred to as vortex-generator and vortex-breaker modes, respectively. This classification implies that one of the processes driving the formation and destruction of major vortical structures in between the buildings is the interaction between low- and high-frequency structures. The high energy revealed by the POD for the vortex-breaker modes points to this destruction process as the mechanism driving the flow dynamics. Furthermore, the results obtained with the STKD method show how the generating- and breaking-mechanisms \rev{originated} along with the streamwise and spanwise directions.
\end{abstract}
\end{frontmatter}

\section{Introduction}

The study of the flow around building-like obstacles has been extensively addressed in the literature~\cite{Hunt1978,Oke1988,Zajic2011} due to its implications in urban-environment phenomena, i.e. pollutant dispersion, air quality, and heat propagation. The very high levels of air pollution \rev{to which} the vast majority of the urban population is exposed are undoubtedly related to \rev{myriad health} issues~\cite{Heaviside2016}. The hunt is for predictive models capable of accurately reproducing the pollutant and thermal distributions within urban environments. Some of these models have already been introduced by the European Union (EU)~\cite{EUUrbWorld}. However, their inability to provide the spatio-temporal accuracy required to model pollutant dispersion through urban environments forces researchers to improve those methods to ensure urban sustainability. For instance, to establish a proper action plan to alleviate the associated adverse consequences, several studies~\cite{Hunt1978,Oke1988,Zajic2011, Becker2002, zhu2017, Bourgeois2012} have focused their efforts on analyzing the spatio-temporal structures of the flow. The main point is to identify the three-dimensional flow regions responsible for the pollutant dispersion within a given urban geometry. Therefore, \rev{this study aims} to apply recently-developed tools from system dynamics, notably higher-order dynamic mode decomposition (HODMD), to turbulent flows within urban environments to understand how different city configurations influence the mechanisms leading the flow dynamics.

\begin{figure}
    \centering
    \includegraphics[width=0.7\textwidth]{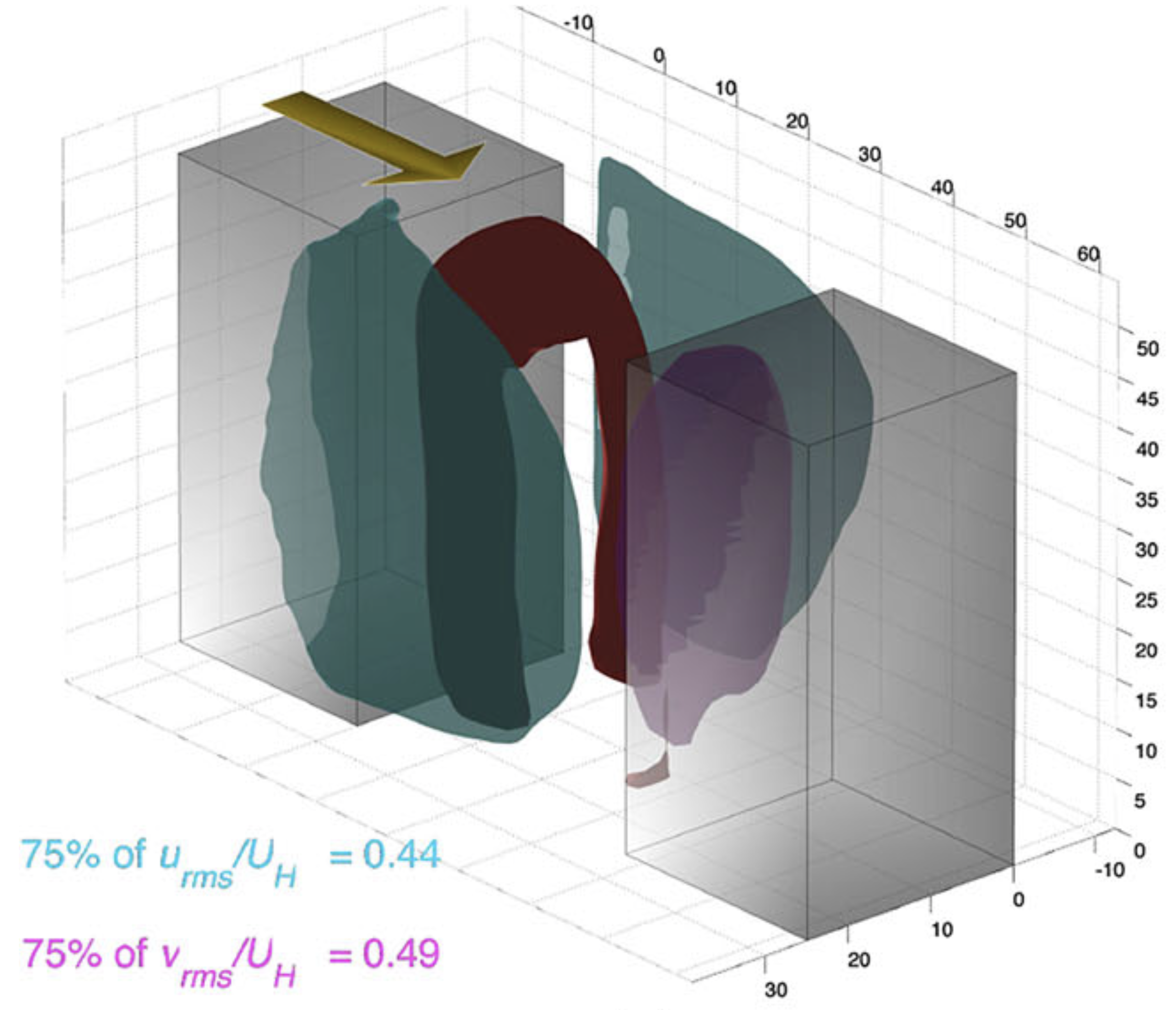}
    \caption{\rev{Arch vortex (red) downstream of the first building represented by isosurfaces of $\Gamma_1=0.4$ for an angle of incidence $AOI=0^\circ$. High-turbulence intensity regions represented by isosurfaces of the root-mean-squared streamwise (green) and spanwise (purple) velocity fluctuations, with threshold values equal to $75\%$ of the maxima in each field (maxima reported in the panel). Figure reproduced from Ref.~\cite{Monnier2018} with permission of the publisher (Springer Nature).}}
    \label{fig: Experimental data}
\end{figure}

The large number of spatio-temporal features present in the high-dimensional nonlinear system of a turbulent flow complicates the analysis. Nonetheless, the fact that physical-flow features are shared across a wide variety of flows suggests that they may be used to describe the dynamics of such a flow. Many experimental~\cite{Hunt1978, Becker2002, zhu2017, Oertel1990, zdravkovich1997,Luo2003,Luo2007,wang_zhou_2009,Bourgeois2011, Monnier2018} and numerical~\cite{zdravkovich1997,Sohankar1999,Saha2003,Vinuesa2015} studies have focused on the flow around a wall-mounted cylinder of different aspect ratios, which is highly three-dimensional~\cite{Becker2002,zhu2017}. Hunt et al.~\cite{Hunt1978} performed one of the first experimental analyses \rev{to} examine the general pattern of the streamlines of the flow around a single wall-mounted bluff obstacle. They proved the absence of a closed surface, i.e. a separation bubble or a cavity, in the wake of the obstacle due to the interaction of four different vortical structures: (I) the horseshoe vortex formed around the obstacle, (II) the roof vortex and (III) the vortices on the obstacles sides, both having a strong interaction with the wake, which yield the formation of the so-called (IV) arch vortex downstream the obstacle. The latter \rev{is depicted in Fig.~\ref{fig: Experimental data} and} consists of two spanwise vortical legs on each side of the obstacle with rotation in the vertical axis and a roof, where the flow rotates in the spanwise direction. These vortices are a consequence of the interaction of the outer flow within the urban-canopy layer. Understanding their underlying physics is essential for developing strategies to reduce pollution dispersion and perform pedestrian-comfort assessment.

The apparent complexity of urban-based environments leads to more intricate physics due to the interaction of flow structures around individual buildings. Oke~\cite{Oke1988} provided an analysis of the resultant flow regimes as a function of the geometrical parameters that define an urban model. Interestingly, the author discovered that the street width was the critical parameter in establishing the flow regimes~\cite{Oke1988}: in the case of narrow streets, the flow above the canopy can barely reach down to the street (skimming flow), and only one vortex can be seen between the obstacles; gradually broader streets lead to the wake-interference regime first and then to the isolated-roughness flow, which exhibits much more contact with the flow above the roofline. Meinders~\cite{Meinders1998} further examined this classification by analyzing the interaction of flow patterns around wall-mounted rectangular obstacles with different spacing ratios in the streamwise direction. The separated shear layer from the first obstacle reattached on the windward side of the downstream obstacle for the lowest separation, resulting in an inter-obstacle area with an arc-shaped vortex confined by the side flow~\cite{Meinders1998}. With larger separation ratios (isolated-roughness regime), flow reattachment occurs in the region between the obstacles, from which a horseshoe vortex emerges around the downstream obstacle. This results in similar flow patterns for both obstacles, but with lower intensity in the downstream block due to the flow disruption of the upstream block~\cite{Meinders1998}.

\rev{A wide range of criteria has been developed to identify these vortical structures}. Monnier et al.~\cite{Monnier2018} aimed at identifying the main flow patterns present in the wind-tunnel flow around the geometry of the Mock Urban Setting Test (MUST) experiment using different criteria. They started evaluating the vorticity components, namely the wall-normal and spanwise mean components, $\hat{\omega}_y$ and $\hat{\omega}_z$, respectively, to identify the location of the arch vortex. Using the modulus of the spatially-averaged vorticity vector allowed them to define a local threshold to properly characterize the influence of the angle of incidence (AOI) on this vortical structure, extracting similar conclusions to those of Becker et al.~\cite{Becker2002}. They also employed some popular methods for vortex identification, based on the second invariant of the velocity gradient tensor, i.e., the Q-criterion~\cite{Hunt1978} and the $\lambda_2$ criterion~\cite{jeong_hussain_1995}. However, they improved the identification of large-scale vortical structures using the normalized angular momentum technique $\Gamma_1$, introduced by Sousa~\cite{Sousa2002} to locate the center of vortical structures downstream of a single cuboid obstacle. This method allowed the authors to describe the relationship between the arch vortex and high-turbulence areas. \rev{As shown in Fig.~\ref{fig: Experimental data}}, Monnier et al.~\cite{Monnier2018} concluded that the arch vortex is located between high-turbulence areas. They consist of two regions of significant streamwise velocity fluctuations on both sides of the obstacles due to the separation of the shear layer and a high spanwise velocity fluctuating region along the windward face of the downstream obstacle. This experimental study led to relevant conclusions in analyzing coherent structures in a more realistic urban model.

Here, we focus on Oke's classification~\cite{Oke1988} to extract through data-driven procedures the key dominant patterns present in the three-dimensional instantaneous fields of the flow through urban environments with different separation ratios. In this regard, we explain the origin and evolution of the various three-dimensional topological patterns that precede the formation of the well-known flow structures found in these geometries: the horseshoe vortex, the roof vortex, the vortices of the obstacles sides, and the arch vortex. For the first time, several modal-decomposition techniques are used to identify how the previous vortices are related to the physical mechanisms driving the \rev{system's dynamics}, shedding light on new possibilities for future urban-flow control research. First, we use proper-orthogonal decomposition (POD)~\cite{Lumley1967} to identify those spatial modes energetically more relevant to the system \rev{and} their associated time coefficients. We compare them with the results obtained using a recently-developed higher-order variant of dynamic-mode decomposition (DMD)~\cite{Schmid2010}, named HODMD~\cite{LeClainche2017b}. Via this novel nonlinear dynamic mode decomposition approach, we can analyze the dynamics of a highly complex turbulent flow~\cite{LeClainche2017,LeClainche2017b,LeClainche2017b,LeClainche2017c,LECLAINCHE2017d,LeClainche2018b}, cleaning noisy artifacts and small amplitude modes from data. Recently, Amor et al.~\cite{Amor2020} showed the potential of HODMD to understand the complicated physics of the wake in a wall-mounted square cylinder, which decomposes spatio-temporal data into a group of modes orthogonal in time, representing the leading flow dynamics~\cite{LeClainche2017b}. Balanced POD (BPOD)~\citep{bpod2005}, spectral POD (SPOD)~\citep{spod2018} and spatio-temporal Koopman decomposition (STKD)~\citep{Clainche2018} are other successful variants of POD and DMD for analysis of turbulent flows. It is notorious the similarities between HODMD and SPOD algorithms. Both methods combine a sliding window process with singular-value decomposition (SVD) to reduce the data dimensionality, selecting the most relevant features of the flow. The successful application of these methods for the analysis of turbulent flows has been presented in several works (e.g., Refs.~\citep{LeClainche2020,clainche2022,spod2018,Schmid2010}). Although both techniques are suitable \rev{for identifying} the main dynamics of the flow in the cases studied in the present article, we will use HODMD. The advantage of this method is that it is automatic. In other words, HODMD identifies the main frequencies driving the flow dynamics without needing previous knowledge about the physics of the case studied~\citep{Mendez2021}. Also, the method provides a complete frequency spectrum driving the flow. Hence it is possible to establish connections between some of the highest-amplitude modes modeling the complex flow dynamics.

Following the introduction to urban flows, the present work provides a general overview of the performed numerical simulations in \S\,\ref{sec: Numerical simulations}. A summary of the mathematical concepts behind the modal-decomposition techniques used to characterize the flow structures over the numerical simulation data is addressed during \S\,\ref{sec: Methodology}. \rev{The main vortices and structures found in the mean flow are presented in \S\,\ref{sec: flow structures}}. The mechanisms driving the flow dynamics within urban environments, which \rev{results from} of the application of different data-driven tools, are investigated in \S\,\ref{sec: Results}. \rev{These structures are further examined in \S\,\ref{sec: POD} using the results obtained from the application of POD.} Finally, a summary of the main conclusions of the project is provided in \S\,\ref{sec: Conclusions}, and the justification of the selected modes is performed in \ref{Appendix: Calibration} through the calibration process of the methods. A review of the formation and destruction mechanisms of arch vortices in urban flows has already been addressed in a companion paper~\cite{laz22}. Here, a detailed analysis of the modal-decomposition techniques, which shed light on the mechanisms driving the flow dynamics, will be addressed with an overview of the high-order numerical simulations carried out to perform the present analysis.

\section{Numerical simulations} \label{sec: Numerical simulations}

The high-order spectral-element code Nek5000~\cite{Nek5000} was used to solve the incompressible Navier--Stokes equations governing the flow in the cases under consideration. Based on the spectral-element method (SEM) of Patera~\cite{PATERA1984}, Nek5000 exhibits both geometrical flexibility and the accuracy of the high-order spectral methods~\cite{Vinuesa2015,Hoyas2006,SIMENS2009, NEGI2018,Noorani2016,VINUESA2021,tanarro_vinuesa_schlatter_2020}. \rev{Due to the flow complexity in urban environments, high-order methods need to be used to resolve all the relevant flow structures properly}. In this database, we use a \textit{well-resolved} large-eddy simulation (LES), the resolution of which is close to that of a direct numerical simulation (DNS)~\cite{NEGI2018}. This code has been extensively used for high-fidelity simulations of complex turbulent flows, see Refs.~\cite{Noorani2016, VINUESA2021, tanarro_vinuesa_schlatter_2020}. \rev{The main focus of the present section is on the parameters with important implications in the modal decompositions}. Additional details on the numerical scheme, employed resolution, and flow statistics can be found in Ref.~\cite{marco2022}.

\begin{figure}
    \centering
    \includegraphics[width=\textwidth]{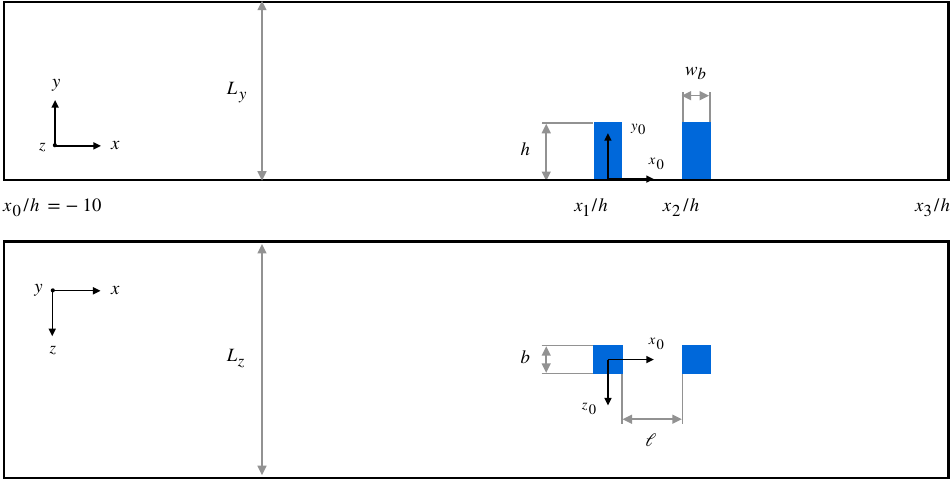}
    \caption{Schematic representation of the numerical domain, where $L_y=3h$ and $L_z=4h$. The flow is from left to right. (Top) and (bottom) show side and top views, respectively.}
    \label{fig: numerical domain}
\end{figure}

The geometrical domain comprises two wall-mounted obstacles, as depicted in Fig.~\ref{fig: numerical domain}. The size of the computational box dimension varies according to the separation of the obstacles. While the wall-normal and spanwise directions remain the same for the three cases, the streamwise length changes proportionally to the separation $\ell$, which modifies the computational cost of the numerical simulation associated with each case. \alvaro{The variation of this parameter according to the flow regime is depicted in Table \ref{tab: Numerical Simulation}.} The obstacles are then defined by the height $h$, length $w_b$ and width $b$. \alvaro{The length-to-height $w_b/h$ and width-to-height $b/h$ ratios correspond to 0.5 for both obstacles}. All dimensions are normalized with the height of the obstacle $h$. The velocity field is given by $\bm{v}(x,y,z,t)$, where $x, y,$ and $z$ are the streamwise, wall-normal, and spanwise directions, respectively, and $t$ is time. Every velocity is normalized with the free stream velocity. The components of the velocity are $\bm{v}=(u,v,w)$, which denote the streamwise, wall-normal, and spanwise components, respectively. Using Reynolds decomposition, $\bm{v}$ is defined as $\bm{v} = V + \tilde{\bm{v}}$, where $V=\overline{\bm{v}}$ is the average in time and $\bm{\tilde{v}}$ is the turbulent fluctuation. Primes are reserved for intensities $\bm{v'}=\overline{\tilde{\bm{v}}^2}^{1/2}$. 

\begin{table}[]
\resizebox{\textwidth}{!}{%
\begin{tabular}{lccccccc}
\hline
\textbf{Flow regime} & Case code & $L_x/h$ & $\ell/h$ & Grid points    & $\Delta T$ & $\Delta t$ & $N_f$ \\ \hline
Skimming flow                         & SF        & 16      & 1        & $6\times 10^6$ & 78.73      & 0.35       & 225   \\
Wake interference                     & WI        & 17      & 2        & $7\times 10^6$ & 64.20      & 0.3        & 215   \\
Isolated roughness                    & IR        & 21      & 4        & $8\times 10^6$ & 62.30      & 0.7        & 90    \\ \hline
\end{tabular}%
}
\caption{Geometrical and temporal parameters of the three flow cases. The reported number of grid points corresponds to the spectrally interpolated mesh used to perform the modal decompositions. The parameter $\ell/h$ refers to the separation-to-height ratio between the obstacles in the streamwise direction. $N_f$ is the number of fields used in the decompositions, $\Delta T$ denotes the time-span and $\Delta t$ refers to the time-step between snapshots. The averaging periods to obtain turbulence statistics follow 40 convective time units, which are discarded to avoid initial transients. All the averaging periods correspond to over 13 eddy-turnover times, based on the $u_{\tau}$ and $h$ values of the turbulent boundary layer (TBL) at $x/h=-2$.}
\label{tab: Numerical Simulation}
\end{table}

As \rev{an} inflow condition, a numerically-tripped~\cite{VINUESA201886, Vinuesa2017} laminar Blasius profile allows the flow for undergoing a rapid transition to turbulence without needing to accelerate the flow before reaching the obstacles. This numerical tripping consists of a weak wall-normal volume randomly added in the forcing terms of the incompressible Navier--Stokes equations to create flow disturbances, thus inducing turbulence. The inflow is located at $x/h = -10$, \rev{and} the tripping force is applied at $x/h=-9$, allowing the boundary layer \rev{to develop} in the region upstream the obstacles, i.e., $-8\leq x/h \leq -1$. In this region, both $z$-averaged friction and momentum-thickness Reynolds numbers, i.e. ${Re}_\tau$ and ${Re}_\theta$ respectively, increase in the streamwise direction, \alvaro{reaching ${Re}_\tau\simeq175$ and ${Re}_\theta\simeq450$ upstream the obstacle at $x/h= -2$, which corresponds to fully-turbulent conditions. The adverse pressure gradient induced by the obstacles leads to an increase of the Rota-Clause pressure-gradient parameter and a decrease in the skin-friction coefficient, which are $\beta = 0.6$ and $C_f = 4.6\times 10^{-3}$ at $x/h= -2$, respectively. In this same location, the boundary-layer thickness evaluated at the $99\%$ of the freestream velocity, $\delta_{99}$, and the shape factor $H$, i.e., the displacement thickness to momentum thickness ratio, are $\delta_{99}=0.38$ and $H=1.62$~\cite{marco2022}}. The stabilized outflow condition developed by Dong et al.~\cite{DONG2014} is used as an outflow condition. At the upper part of the domain, a combination of outflow and Dirichlet conditions is used to simulate an open-air urban environment: a zero-stress condition is applied in the wall-normal direction and a Dirichlet condition in the other two directions~\cite{Nek5000}. Finally, periodicity is applied in the spanwise direction. A smooth-wall (including the no-slip and no-penetration) condition is applied to the bottom plane of the domain and the surfaces of the obstacles.

We consider a spectral-element mesh with an eight-point Gauss--Lobatto--Legendre (GLL) quadrature in each element to solve the scale disparity of the flow. The mesh is refined in the near-obstacle area to increase resolution, which has a direct impact on flow statistics. Following the criteria of Negi et al.~\cite{NEGI2018}, the mesh employed in this work satisfies all the resolution criteria to be considered a \textit{well-resolved} LES. \alvaro{Furthermore, the study focuses only on neutral stability conditions, where no buoyancy effects are simulated~\cite{garratt1994}.}

\alvaro{As we focus on the flow near the obstacles, the following region is extracted from the computational domain: $0\leq y/h \leq 2$ and $-1.5\leq z/h \leq1.5$. For the streamwise direction, we use $-1\leq x/h \leq5$ for the SF regime, $-1\leq x/h \leq7$ for WI, and $-1\leq x/h \leq11$ for IR, owing to the change in the separation between the obstacles. Using this reduced domain, we consider 225, 215, and 90 three-dimensional instantaneous fields of the three components of the velocity to perform the modal decompositions on the SF, WI, and IR flow regimes, respectively. Note that additional analyses conducted with a larger number of snapshots did not yield significant differences in the large-scale structures identified by the methods; hence, results are considered to be converged with the previous sets of snapshots. The previous fields were spectrally interpolated from the original SEM mesh to another one with a coarser resolution}. The analyzed database has the temporal parameters gathered in Table \ref{tab: Numerical Simulation}. Note that this information is critical for the analysis of the spatio-temporal structures of the flow since they define the dynamical behavior of the system, which is closely related to the time span and time step of the snapshots to be analyzed. \alvaro{On this account, every database is obtained over a time span of the same order of magnitude, which is sufficient to accurately capture the low-frequency mechanisms occurring in the flow.} All the introduced parameters are expressed in convective time units, i.e. a ratio between a characteristic length and a velocity. In the present \rev{work}, time is obtained from the freestream velocity $U_\infty$ and the height of the obstacle, $h$. In all cases, the Reynolds number is set to $Re_h=10,000$.

\section{Methodology for modal decomposition} \label{sec: Methodology}

\subsection{Proper-orthogonal decomposition (POD)}
The proper orthogonal decomposition (POD) is a modal-decomposition technique, introduced \rev{in fluid mechanics} by Lumley~\cite{Lumley1967}, which aims at extracting coherent patterns from a given flow field. Thus, the objective of the POD algorithm is to decompose a set of data of a given field variable into a minimal number of modes (basis functions) that capture as much energy as possible. This process implies that POD modes are optimal in minimizing the mean-square error between the signal and its reconstructed representation. For instance, if the field variable to be examined is the velocity, the modes representing such variable are optimal to capture the kinetic energy of the flow field. This low-dimensional latent space provided by the POD modes is attractive for interpreting the most energetic and dominant patterns within a given flow field. Let us consider a vector field $\bm{q}\left(\bm{\xi},t\right)$, which may represent the velocity or the vorticity field depending on a spatial vector $\bm{\xi}$ and time. In fluid-flow applications, subtracting the temporal mean $\bm{\bar{q}}\left(\bm{\xi}\right)$ allows for the analysis of the unsteady component of the field variable:
\begin{equation}
    \bm x(t) = \bm{q}\left(\bm{\xi},t\right) - \bm{\bar{q}}\left(\bm{\xi}\right),\quad\quad t = t_1,t_2,\dots,t_k
\end{equation}
\noindent where $\bm x(t)$ represents the fluctuating component of the vector data with its temporal mean removed. This representation emphasizes that the data vector $\bm x(t)$ is considered as a collection of snapshots at different time instants $t_k$. If the $m$ snapshots are then stacked into a matrix \rev{form}, we obtain the so-called snapshot matrix $\bm X$:
\begin{equation}
    \bm X = \left[\bm x(t_1), \bm x(t_2),\dots,\bm x(t_m)\right] \in \mathbb{R}^{J\times K},
\end{equation}
\noindent where $J$ represents the number of points in $x$, $y$ and $z$. The objective of the POD analysis is to find the optimal basis to represent the given set of data $\bm x(t)$. This can be solved finding the eigenvectors $\bm{\Phi}_j$ and the eigenvalues $\lambda_j$ from:
\begin{equation} \label{eq: POD basic eq}
    \bm C \bm{\Phi}_j = \lambda_j \bm{\Phi}_j, \quad\quad \bm{\Phi}_j \in \mathbb{R}^{J}, \quad\quad \lambda_1\geq\dots\geq\lambda_N\geq0,
\end{equation}
\noindent where $\bm C$ states for the covariance matrix of the input data, defined as
\begin{equation}
    \bm C = \sum_{i=1}^{K} \bm x\left(t_i\right)\bm x^\text{T}\left(t_i\right) = \bm X \bm X ^\text{T} \in \mathbb{R}^{J\times J}.
\end{equation}
The size of this matrix depends on the spatial degrees of freedom of the problem. In the case of fluid flows, this value is usually large since it equals the number of grid points times the variables to be considered. The POD modes are derived from the eigenvectors of Eq.~(\ref{eq: POD basic eq}), with the eigenvalues reflecting how well each eigenvector $\bm{\Phi}_j$ represents the original data in \rev{a least-squares optimal sense, i.e. it offers a way to find lower-dimensional linear approximations of a given data set}~\cite{Lumley1967}. This enables a hierarchy of modes in terms of captured energy, which improves understanding of the most prominent patterns, e.g. in a specific flow field.

Another method for computing the POD algorithm is based on the singular-value decomposition (SVD)~\cite{Sirovich1987}, which can be applied directly on the snapshot matrix $\bm X$ to obtain the left $\bm{\Phi}$ and right $\bm \Psi$ singular vectors as
\begin{equation}
    \bm X = \bm\Phi \bm\Sigma \bm\Psi ^\text{T},
\end{equation}
\noindent where $\bm\Phi \in \mathbb{R}^{J\times J}$, $\bm\Psi \in \mathbb{R}^{K\times K}$ and $\bm\Sigma \in \mathbb{R}^{J\times K}$. The matrix $\bm \Sigma$ contains the singular values $\left(\sigma_1,\sigma_2,\dots,\sigma_N\right)$ along its diagonal, which relates to the eigenvalues as $\sigma_j^2 = \lambda_j$. Moreover, the left and right singular vectors correspond to the eigenvectors of matrices $\bm X \bm X ^\text{T}$ and $\bm X ^\text{T}\bm X$, respectively. Therefore, the SVD can be seen as a rectangular-matrix decomposition technique capable of computing the POD modes.

\subsection{Higher order dynamic mode decomposition (HODMD)} \label{sec: HODMD methodology}
Aiming at identifying the spatio-temporal coherent patterns present in high-dimensional flow data, Schmid~\cite{Schmid2010} developed a data-driven tool \rev{that} retrieved the spatially-correlated structures with similar behavior in time. This methodology, known as dynamic-mode decomposition (DMD), provides not only a reduction in dimension concerning a reduced set of modes \rev{that} best reproduce the input flow field, but also a model for the interaction of those modes in time.

The method decomposes the vector field data $\bm v\left(\textbf{x},t\right)$ as an expansion of $M$ Fourier-type modes:
\begin{equation} \label{eq: Fourier modes}
    \bm v\left(\textbf{x},t\right) \simeq \sum_{m=1}^M a_m {\bm u}_m\left(\textbf{x}\right)e^{\left(\delta_m+i{\omega}_m\right)t_k},
\end{equation}
\noindent for $k=1,...,K$, where ${\bm u}_m$ represents the DMD modes weighted by an amplitude $a_m$, ${\omega}_m$, their associated frequencies and $\delta_m$, their associated growth rates, which symbolize the temporal growth or decay of the ${\bm u}_m$ modes in time.

The standard DMD algorithm assumes a linear relationship of two consecutive snapshot matrices using the linear Koopman operator \textbf{R}. To this end, a general snapshot matrix $\textbf{V}_{k_1}^{k_2}$ can be defined for $k_1<k_2$ so that its columns represent the snapshots varying equidistantly between $k_1$ and $k_2$, namely:
\begin{equation}
    \textbf{V}_{k_1}^{k_2} = \left[{\bm v}_{k_1}, {\bm v}_{k_1+1}, ..., {\bm v}_{k_2}\right].
\end{equation}
Therefore, using the previous nomenclature, the standard DMD can be defined based on the Koopman operator as
\begin{equation}
    \textbf{V}_{2}^{K} \simeq \textbf{R} \textbf{V}_{1}^{K-1},
\end{equation}
\noindent where $\textbf{V}_{2}^{K}$ and $\textbf{V}_{1}^{K-1}$ represent here the second to last snapshots and the first to the second last snapshots of the data matrix, respectively. Recalling Eq.~(\ref{eq: Fourier modes}), this equation might be seen as the simplest equation exhibiting such behavior~\cite{Vega2020Book}. The Koopman matrix \textbf{R}, which is independent of $k$, contains the dynamical information of the system. Recently, Le Clainche \& Vega~\cite{Vega2020Book} extended the DMD method for the analysis of various types of flows, e.g. turbulent, multi-scale or transitional flows and noisy experimental data. Based on Takens' delayed-embedded theorem~\cite{Vega2020Book}, the higher order dynamic mode decomposition (HODMD) relates $d$ time-delayed snapshots using higher-order Koopman assumption defined as
\begin{equation} \label{eq: HO Koopman General}
    \textbf{V}_{d+1}^{K} \simeq \textbf{R}_1 \textbf{V}_{1}^{K-d} + \textbf{R}_2 \textbf{V}_{2}^{K-\left(d-1\right)} + ... + \textbf{R}_d \textbf{V}_{d}^{K-1},
\end{equation}
which relates each flow field with the $d$ subsequent fields. The HODMD algorithm can be encompassed into \rev{two} main steps.

\subsubsection{Step 1: Dimension reduction} \label{sec: DMD Step 1}

First of all, the SVD technique is employed to reduce spatial redundancy and filter out noise caused by numerical or experimental errors. The truncated SVD allows for the reduction of the original snapshot data into a series of linearly independent vectors of dimension $N$ (where $N<J$ is the spatial complexity), based on a certain tolerance $\varepsilon_{\text{SVD}}$:
\begin{equation} \label{eq: SVD decomposition DMD}
    \textbf{V}_1^K \simeq \textbf{W}\mathbf{\Sigma} \textbf{T}^\text{T},
\end{equation}
\noindent where $\mathbf{\Sigma}$ includes the singular values $\sigma_1,...,\sigma_N$ and $\textbf{W}^\text{T}\textbf{W}=\textbf{T}^\text{T}\textbf{T}={\bm I}$ are $N\times N$ unitary matrices. Note that the parameter $\varepsilon_{\text{SVD}}$ is tunable based on previous information of the simulation or experimental data, e.g. if the noise level of the snapshots is known in advance, then $\varepsilon_{\text{SVD}}$ may be set to be comparable to that level (see details in Ref.~\cite{LECLAINCHE2017d}). Above all, this parameter determines the number $N$ of SVD retained modes as:
\begin{equation} \label{eq: Dimension reduction Step 1}
    \frac{\sigma_{N+1}}{\sigma_1} \leq \varepsilon_{\text{SVD}}.
\end{equation}
Following the definition in Eq.~(\ref{eq: SVD decomposition DMD}), the reduced snapshot matrix $\hat{\textbf{T}}$ can be defined as:
\begin{equation} 
    \textbf{V}_1^K \simeq \textbf{W}\mathbf{\Sigma} \textbf{T}^\text{T} \equiv \textbf{W} \hat{\textbf{T}}_1^K.
\end{equation}
The dimension of this reduced snapshot matrix is $N\times K$.

\subsubsection{Step 2: The DMD-d algorithm} \label{sec: DMD Step 2}

The higher-order Koopman assumption, defined in Eq.~(\ref{eq: HO Koopman General}), is now applied to the \alvaro{modified} snapshot matrix as:
\begin{equation}
    \hat{\textbf{V}}_{d+1}^{K} \simeq \hat{\textbf{R}}_1 \hat{\textbf{V}}_{1}^{K-d} + \hat{\textbf{R}}_2 \hat{\textbf{V}}_{2}^{K-\left(d-1\right)} + ... + \hat{\textbf{R}}_d \hat{\textbf{V}}_{d}^{K-1},
\end{equation}
\noindent where $\hat{\textbf{R}}_k=\textbf{W}^\text{T}{\textbf{R}}_k\textbf{W}$ is used for $k=1,...,d$. The above equation may be cast in a more generic form by incorporating the modified snapshot matrix $\tilde{\textbf{V}}_1^{k-d+1}$ and the modified Koopman matrix $\tilde{\textbf{R}}$ as:
\begin{equation} \label{eq: general koopman matrix}
    \tilde{\textbf{V}}_2^{K-d+1} = \tilde{\textbf{R}} \tilde{\textbf{V}}_1^{K-d},
\end{equation}
\noindent where the many Koopman operators $\hat{\bm{R}}_1,\dots,\hat{\bm{R}}_K$ are then combined into a single matrix after some computations, from which the eigenvalue problem can be solved to obtain the DMD modes, frequencies and growth rates defining the DMD expansion of Eq.~(\ref{eq: Fourier modes}). Sorted in decreasing order of the mode amplitudes, this expansion is further reduced by removing the modes such that:
\begin{equation} \label{eq: Dimension reduction Step 2}
    a_m/a_1 < \varepsilon_\text{DMD},
\end{equation}
\noindent for $m = 1,\dots,M$, where $\varepsilon_\text{DMD}$ represents a parameter tunable by the user. The number of retained modes, $M$, represents the spectral complexity of the analysis. This complexity, together with the spatial one, determines the performance of the HODMD algorithm, which reduces to the standard DMD when $d=1$. In complex fluid flows, the spatial complexity is usually smaller than the spectral one, $N<M$, where the standard DMD fails, thus requiring the use of the DMD-d algorithm. Furthermore, using the tunable parameters $\varepsilon_\text{SVD}$ and $\varepsilon_\text{DMD}$ enables retaining only the large scales of the input data, which is particularly interesting for complex turbulent flows involving a large number of scales.

\subsection{Spatio-temporal Koopman decomposition (STKD)} \label{sec: STKD methodology}
Spatio-temporal Koopman decomposition (STKD) is an extension of HODMD introduced to identify spatio-temporal structures as an expansion of traveling and standing waves driving the flow dynamics both in the streamwise and spanwise directions. For the streamwise direction, the spatio-temporal modes $\mathbf{u}_{mn_1}$ and growth rates $\nu_{mn_1}$ are defined in the following modal expansion, which reconstruct the original flow field analyzed as:
\begin{equation}
	\bm{v}(x_j,y,z,t_{k})\simeq  
	\sum_{m,n_1=1}^{M,N_1} a_{mn_1}\widehat{\mathbf{u}}_{mn_1}(y,z)e^{(\delta_m+i \omega_m)t_k+(\nu_{mn_1} + i \alpha_{mn_1})x_j},\label{ab002}
\end{equation}
\noindent for $k=1,\ldots,K$ and $j=1,\ldots,J$. It must be emphasized that the spatio-temporal expansion is useful when the data exhibit exponential/oscillatory behavior in both the $x$ coordinate and time. In this case, it is interesting to compare the expansion (\ref{ab002}) with the purely temporal
expansion (\ref{eq: Fourier modes}). \rev{The former} can be easily obtained by simply applying HODMD to the DMD modes in Eq.~(\ref{eq: Fourier modes}), resulting in the following DMD expansion:
\begin{equation}
	\mathbf{u}_m(x_j,y,z)\simeq  
	\sum_{n_1=1}^{N_1} a_{n_1}\widehat{\mathbf{u}}_{mn_1}(y,z)e^{(\nu_{mn_1} + i \alpha_{mn_1})x_j},\label{ab003}
\end{equation}
\noindent for $j=1,\ldots,J$. Eq.~(\ref{ab002}) is obtained by combining this solution with Eq.~(\ref{eq: Fourier modes}), where the spatio-temporal amplitudes are defined as $a_{mn_1}=a_m a_{n_1}$. In a similar way, it is possible to obtain spatio-temporal expansions defined along the spanwise direction as:
\begin{equation}
	\bm{v}(x,y,z_r,t_{k})\simeq  
	\sum_{m,n_1=1}^{M,N_1} a_{mn_1}\bar{\mathbf{u}}_{mn_1}(x,y)e^{(\delta_m+i \omega_m)t_k+(\lambda_{mn_1} + i \beta_{mn_1})z_r},\label{ab004}
\end{equation}
\noindent for $k=1,\ldots,K$ and $r=1,\ldots,R$, where $\lambda_{mn_1}$ and $\beta_{mn_1}$ are the growth rates and wavenumbers related with the spanwise direction. Using this expansion, it is also possible to describe the analyzed data as a group of traveling waves, the phase velocity of which is defined as $c_{mn_1}=\omega_m/\beta_{mn_1}$.  A more detailed description of the method can be found in Refs.~\cite{Clainche2018,LeClainche2018b}. Also, notice that this method has been successfully applied to turbulent non-periodic flows \cite{LeClainche2020,le2022data}.

\rev{Finally, we conclude this section with a brief comparison of the previous techniques for the sake of completeness. While POD decomposes a given field variable into a set of linearly-superposed modes orthogonal in space and capturing as much energy as possible, HODMD offers a more advantageous decomposition when the dynamics of the system are the focus of the study. On the one hand, the dynamical behavior of the variables \rev{is} not considered in the decomposition provided by POD, but is rather included in a set of time-varying coefficients associated with each of the modes. In contrast, HODMD decomposes the spatio-temporal data into a set of Fourier-type modes ranked in terms of their dynamical behavior, i.e. frequencies, amplitudes, and growth rates. Therefore, a combined analysis using HODMD and POD provides insights into the most relevant mechanisms of the flow in terms of dynamics and energy, respectively. Finally, STKD is an extension of HODMD, which identifies spatio-temporal structures as a collection of modes and waves. In this case, the method is particularly useful when the flow dynamics exhibit exponential or oscillatory behaviour in the streamwise and spanwise directions and in time.
}

\section{Mean-flow structures} \label{sec: flow structures}

\rev{In this section, we focus on the evolution of the primary structures characterizing the mean flow for the three flow regimes. These flow regimes are the skimming flow (SF), the wake interference (WI) and the isolated roughness (IR)~\cite{Oke1988}. The change from a cavity-like to a wave-like flow topology is the most noticeable result of increasing the distance between the obstacles~\cite{marco2022,Zhao2021}. They both represent a substantial recirculation zone in the rear part of the upstream obstacle; however, for the cavity-like flow, this recirculation region extends throughout the entire space between the obstacles, whereas for the wave-like topology, it is only restricted to the immediate rear part of the first obstacle~\cite{marco2022}.} Meinders~\cite{Meinders1998,Meinders2002} also extended the work on flow structures around wall-mounted cubes by analyzing the interaction between the obstacles when more than one cuboid was introduced. Using oil-film visualizations, Meinders experimentally analyzed the influence of the separation distance between the obstacles, i.e. $\ell$, on the flow around an in-line tandem disposition of two cubes. It was proved that the separation variance only led to a substantial modification of the mean flow patterns \rev{and, as a consequence, of the interaction of the free stream within the inter-obstacle region.} 

\begin{figure}[t]
     \centering
     \begin{subfigure}[b]{0.32\textwidth}
         \centering
         \includegraphics[width=\textwidth]{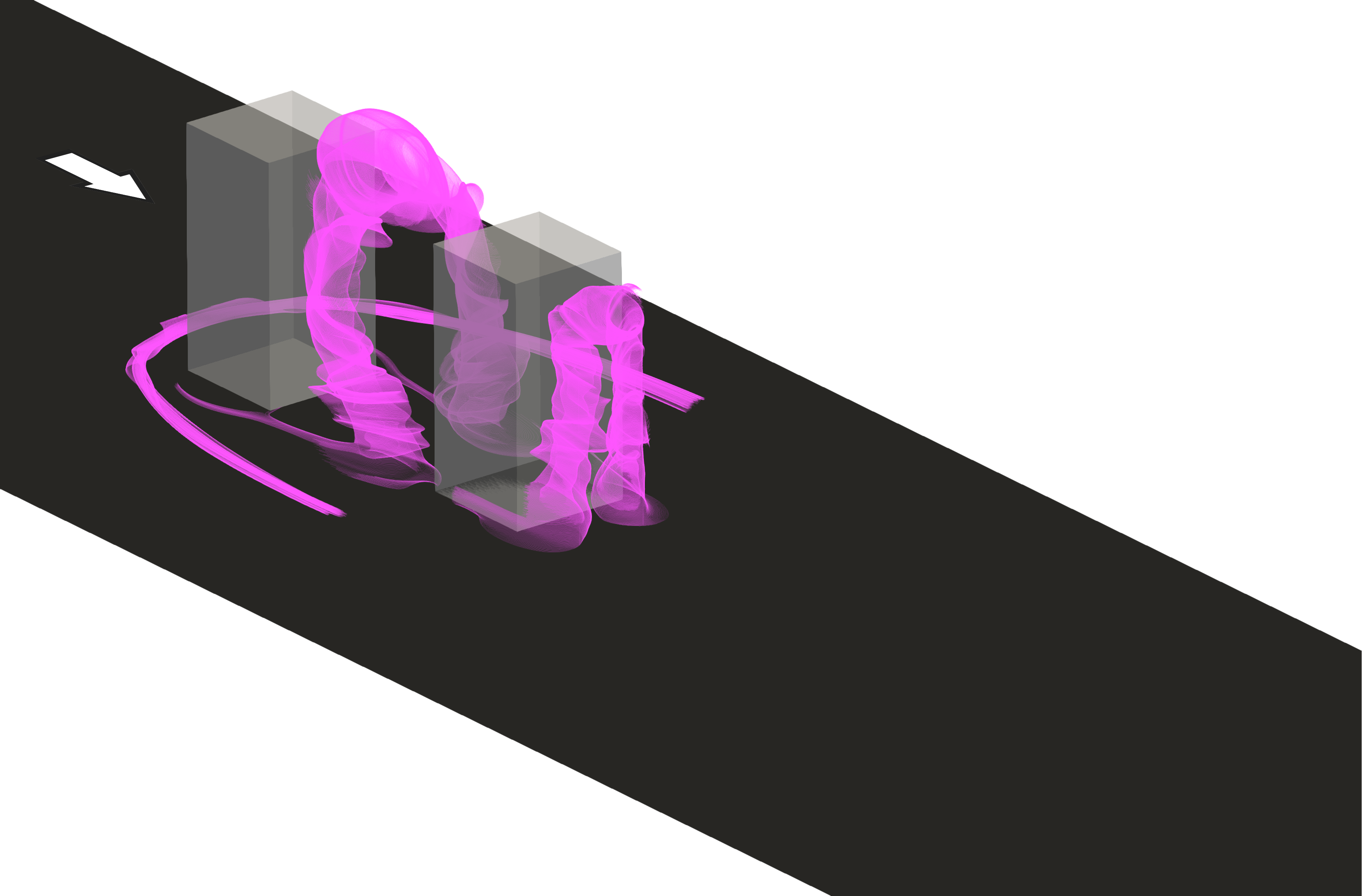}
         \caption{Skimming flow}
     \end{subfigure}
     \begin{subfigure}[b]{0.32\textwidth}
         \centering
         \includegraphics[width=\textwidth]{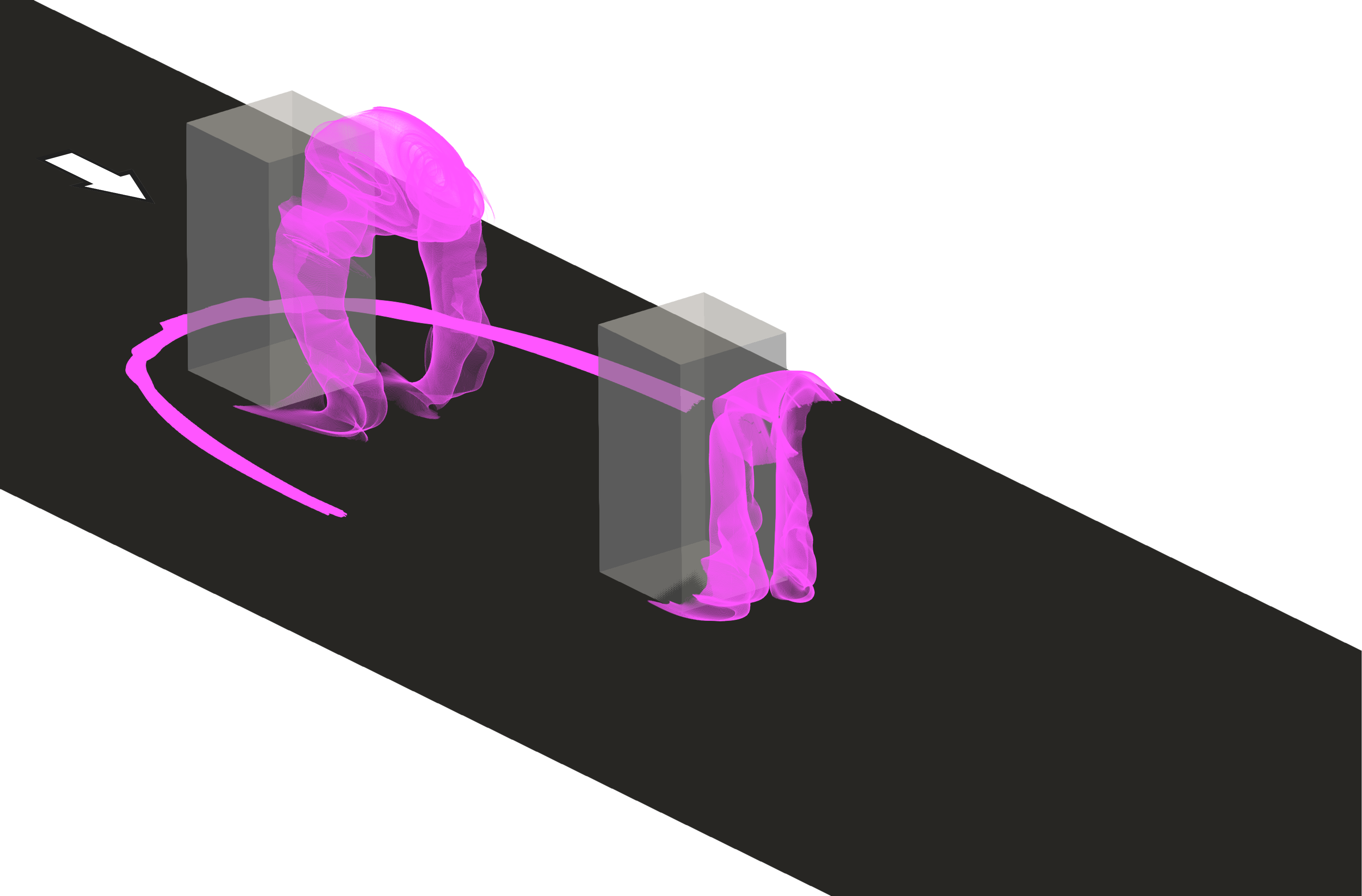}
         \caption{Wake interference}
     \end{subfigure}
     \begin{subfigure}[b]{0.32\textwidth}
         \centering
         \includegraphics[width=\textwidth]{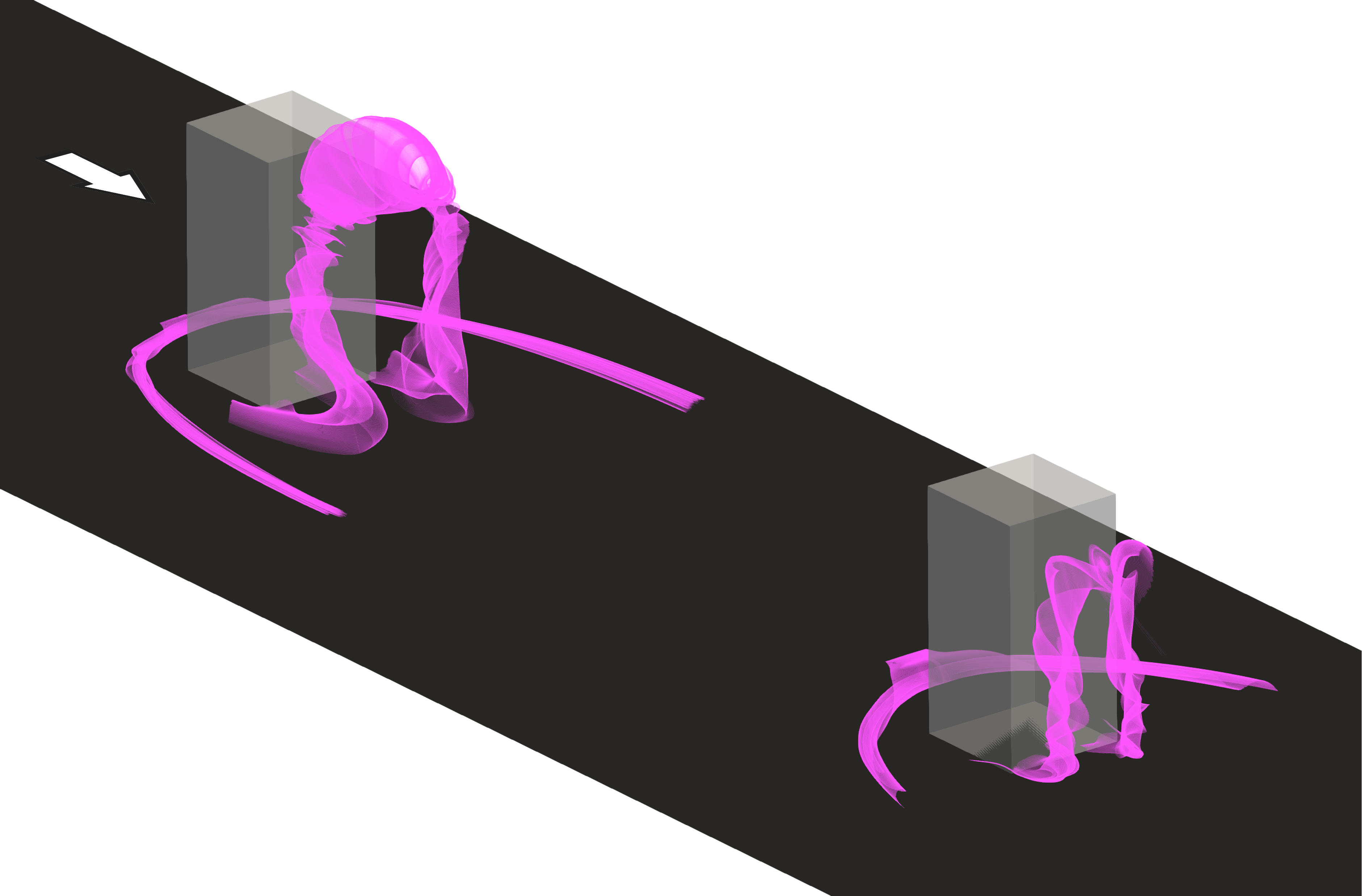}
         \caption{Isolated roughness}
     \end{subfigure}
     \caption{Main vortical structures formed around two wall-mounted obstacles with different separation ratios: (\textbf{a}) $\ell/h=1$, (\textbf{b}) \alvaro{$2$} and (\textbf{c}) \alvaro{$4$}, visualized \rev{employing} streamlines. Note the arch on the leeward side of the obstacles. The arrow indicates flow direction.}
     \label{fig: Arch Vortex}
\end{figure}

In Fig.~\ref{fig: Arch Vortex}, the streamline time-averaged flow patterns for the three flow regimes are depicted. The main conclusions \rev{addressed here} are in good agreement with the experimental results of Meinders~\cite{Meinders1998}. For the lowest separation $\ell/h=1$, i.e., the skimming-flow regime, \rev{the inter-obstacle region is characterized by an arch-shaped vortex. This means that the flow between the obstacles is fully confined by the flow on the sides and the roof, a feature characteristic of cavity-like flows}. In addition, a horseshoe vortex emerges upstream of the windward face of the leading cube, and it is deflected downstream along the sides of both obstacles\rev{, as a result of the low penetration of the flow above and around the sides of the canopy}. For the isolated-roughness regime, i.e., $\ell/h=4$, the \rev{surrounding} flow eventually \rev{interacts with} the inter-obstacle region: a shear layer detached from the sides and top \rev{sharp} edges of the upstream obstacle breaches the inter-obstacle spacing before \rev{reaching the downstream obstacle}. Because of that flow \rev{interaction within the region in between the obstacles}, a second horseshoe vortex emerges around the downstream cube\rev{, which is characteristic of wave-like flows and is not observed for the previous case}. The wake-interference regime, with a separation ratio $\ell/h = 2$, exhibits similar flow patterns to those of the SF and IR regimes. In this case, the arch vortex does not span the \rev{entire region} between the obstacles. \rev{This is characteristic of a transition from a cavity-like to a wake-like flow:} the separation is large enough for the \rev{free stream} to slightly interact with the inter-obstacle region. \rev{Nevertheless, this interaction is not large enough for the flow to exhibit independent structures, like a horseshoe vortex, around the second obstacle.}

Remarkably, in the three flow regimes, the wake behind the second obstacle is relatively similar: a second arch vortex is formed on the leeward side of the downstream obstacle, albeit with a lower intensity. This is mainly due to the flow disruption of the upstream obstacle, which induces a different turbulent-intensity level upstream of the second one~\cite{Oke1988,Meinders2002,Vinuesa2015}. Finally, the reader is referred to Ref.~\cite{marco2022} for a more detailed discussion on the flow statistics \rev{for the three regimes}.

\section{Spatio-temporal structures}\label{sec: Results}

\begin{figure}[t]
    \centering
    \includegraphics[width=\textwidth]{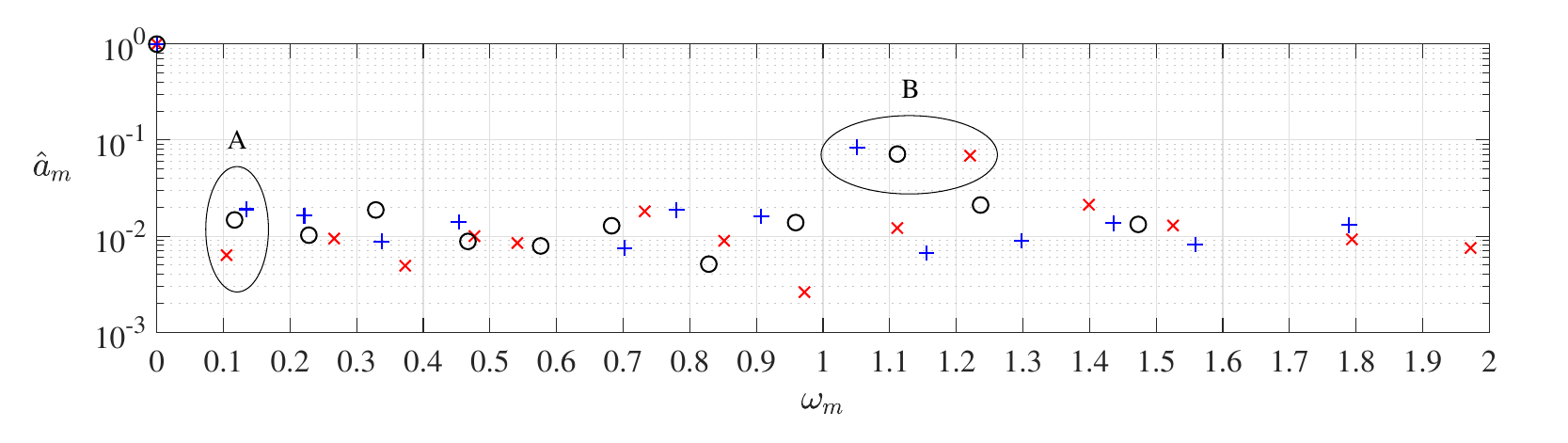}
    \caption{DMD-d modes. Amplitude scaled with its maximum value ($\hat{a}_m=a_m/a_0$) versus frequency ${\omega}_m$ computed for ({red}) skimming flow, ({blue}) wake interference and ({black}) isolated roughness. The modes represented here are the result of a calibration process of the user parameters, from where $\varepsilon_\text{SVD} = \varepsilon_\text{DMD} = 10^{-3}$, $d=20$ for SF ($K=225$ snapshots) and WI ($K=215$ snapshots) and $d=10$ for IR ($K=90$ snapshots) have been selected as reference. Note that the A and B modes highlighted here are examples of vortex-generator and vortex-breaker modes, respectively.}
    \label{fig: DMD spectrum}
\end{figure}

Data-driven modal decompositions are powerful techniques to extract the energetically important features from a given flowfield. Having identified the main flow structures present in the time-averaged fields in \S\,\ref{sec: flow structures}, the analysis \rev{of} the instantaneous fields will allow for the characterization of the main mechanisms driving the flow dynamics. \alvaro{To that end, the results obtained with a highly-efficient tool for the analysis of complex flows, i.e. higher-order DMD, are analyzed in this section. As seen in \S\,\ref{sec: HODMD methodology}, an important characteristic of this method is the higher-order Koopman assumption, which relates each flow field with $d$ subsequent fields. Therefore, a proper selection of this parameter is paramount for the identification of the main mechanisms of the flow, specially when the amount of data is restricted.} In \ref{Appendix: Calibration} we provide a detailed overview of the calibration process of the identification method of the modes. In particular, Fig.~\ref{fig: DMD spectrum} shows the frequency versus amplitude of the different modes computed using DMD-d for the three flow regimes. These results were obtained using the databases specifications from Table \ref{tab: Numerical Simulation}. \alvaro{A dominant mode, i.e. the one with highest amplitude, is located between the frequencies $\omega_m=1$ and $\omega_m=1.3$, while the rest of the modes are subharmonics and harmonics of it. Additionally, another relevant mode is the one with the lowest frequency, $\omega_m=0.1$, since it is the first mode to appear in the spectrum and it is the frequency that drives the periodicity of the main physics.}

\begin{figure}[t]
    \centering
    \includegraphics[width=\textwidth]{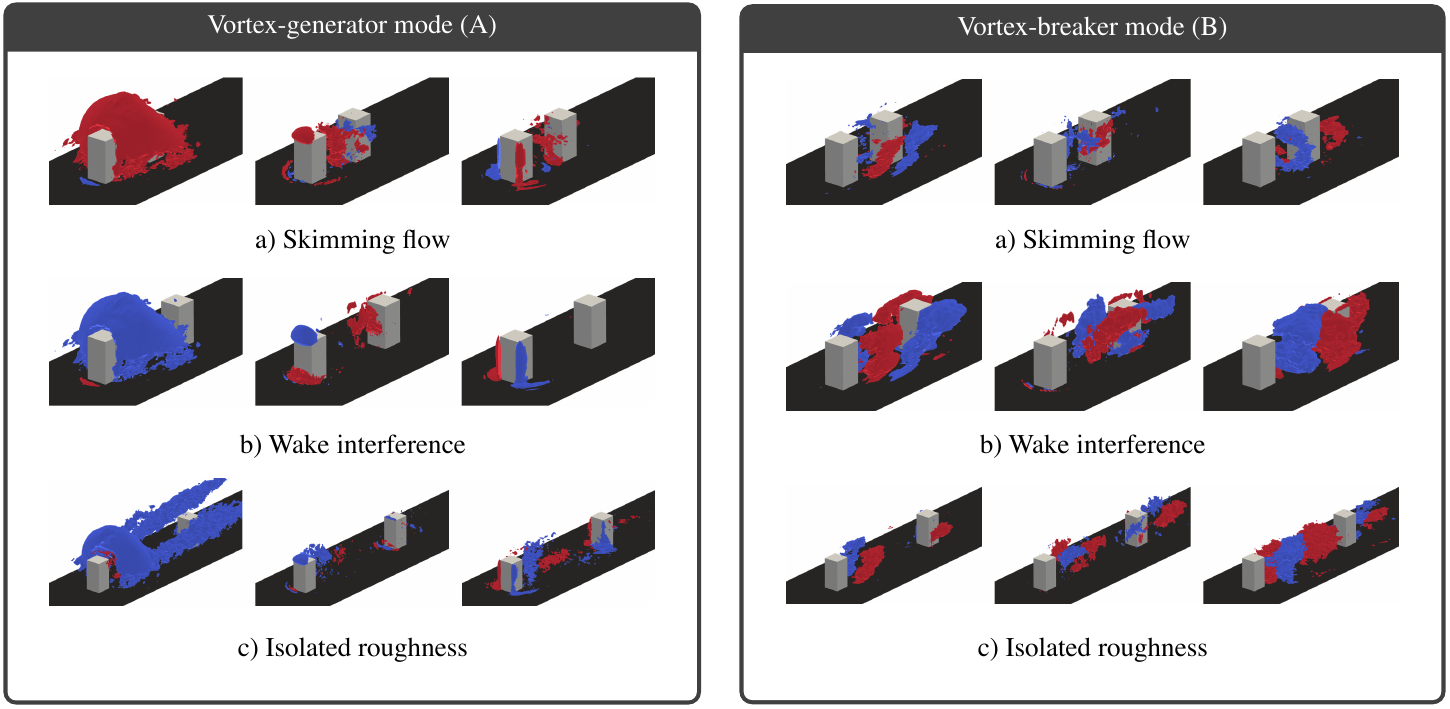}
    \caption{Three-dimensional iso-surfaces of the ({left}) streamwise, ({middle}) wall-normal and ({right}) spanwise velocities of the vortex-generator (A) and vortex-breaker (B) modes highlighted in Fig.~\ref{fig: DMD spectrum}. The left and right panels refer to vortex-generating and vortex-breaking modes, respectively. In each panel, the flow moves from left to right. Velocity values are normalized using the $L_\infty$-norm. The iso-values employed are given by $a\,U_\text{max}$ (blue) and $b\,U_\text{min}$ (red): for the vortex-generator mode, (left) streamwise velocity is represented with $a=0.6$ and $b=0.7$, (middle) wall-normal velocity is depicted using $a=0.6$ and $b=0.5$ and (right) spanwise velocity is represented with $a=b=0.5$; for the vortex-breaker mode, $a=b=0.4$ for all velocity components.}
    \label{fig: HODMD main modes}
\end{figure}

\alvaro{Based on this identification, we highlight two types of modes, which are also shown in Fig.~\ref{fig: HODMD main modes}: vortex-generator (A) modes and vortex-breaker (B) modes. Vortex-generator modes are the ones responsible for the main flow structures and vortices; therefore, this suggests that they could be related to the mechanism that creates the arch and horseshoe vortices. These modes usually appear in the low-frequency area of the spectrum and have a lower amplitude than B modes. For instance, for the WI flow regime, some of these modes appear at the frequencies $\omega_m = 0.11$, $0.22$ and $0.33$. Conversely, the vortex-breaker modes are suggested to be responsible for the destruction mechanisms of the main flow structures and the ones that generate the turbulent wake. As opposed to A modes, B modes appear in the high-frequency region of the domain, i.e. for the ninth harmonic of the lowest frequency onward. \rev{Further studies need to be carried out to study in detail the main mechanism and instabilities connected to the presence of the previous modes}.

To verify this classification of modes in a quantitative way, the modal-assurance criterion (MAC) analysis~\cite{Pastor2012,Mendez2021} was used. Given two complex modal vectors $\mathbf{u}_i$ and $\mathbf{u}_j$, the MAC value represents the normalized dot product of the modal vectors at common points, namely:
\begin{equation}
    \rm{MAC} = \frac{\left< \mathbf{u}_i, \mathbf{u}_j \right>}{ \lVert \mathbf{u}_i \rVert\, \lVert \mathbf{u}_j \rVert}.
\end{equation}
\rev{If the mode's shapes are identical, the MAC will have a value of 1. In the present study, modes with $\rm{MAC}>0.8$ are considered similar}. Note that in this study $\rm{MAC}<1$, since the large range of scales found within turbulent flows play a role. Note as well that the relative error made in the calculations when comparing the reconstructed and original fields always remains fenced in the set of tolerances used. Different calculations with these tolerances, defined in Eqs.~(\ref{eq: Dimension reduction Step 1}) and~(\ref{eq: Dimension reduction Step 2}), were made during the calibration process, see \ref{Appendix: Calibration}. The goal of this study, however, is to identify some of the large-scale structures that provide a broad description of the fundamental patterns driving the flow, rather than to build any accurate reduced-order models based on the physical knowledge of the flow. As a result, the relative error will not be examined further in this study.}



\alvaro{It has been demonstrated that areas with substantial recirculation cause an increase in the concentration of passive scalars~\cite{zhu2017}, \rev{i.e. the diffusive contaminants present in a fluid flow that have no dynamical effect on the fluid motion.} As a result, according to Monnier et al.~\cite{Monnier2018}, A modes, which are associated with these prominent recirculating zones, may be linked to areas of high pollution concentration. Similarly, since B modes are related to the processes that cause these coherent structures to shatter, it is possible to hypothesize that they are associated with areas of lower pollution concentration. In particular, the arch vortex is a recirculating feature formed in the region in between the building-like obstacles. Therefore, further understanding of these flow patterns could allow for the development of some control strategies that might reduce the presence of A modes in urban areas, hence, reducing pollutant concentration. However, further research needs to be carried out to identify these types of modes in databases modeling multi-phase flows, which remains as a topic for future research.} 

Fig.~\ref{fig: HODMD main modes} shows a three-dimensional view of the main HODMD modes presented in Fig.~\ref{fig: DMD spectrum} as a function of the separation ratio between the obstacles. Fig.~\ref{fig: HODMD main modes} (left) corresponds to the vortex-generator low-frequency mode, mode A, whereas the vortex-breaker high-frequency mode, mode B, is depicted in Fig.~\ref{fig: HODMD main modes} (right). These results are further discussed in the following sections.

\subsection{Generation process of the main vortices}

\begin{figure}[t]
     \centering
     \begin{subfigure}[b]{\textwidth}
         \centering
         \includegraphics[width=\textwidth]{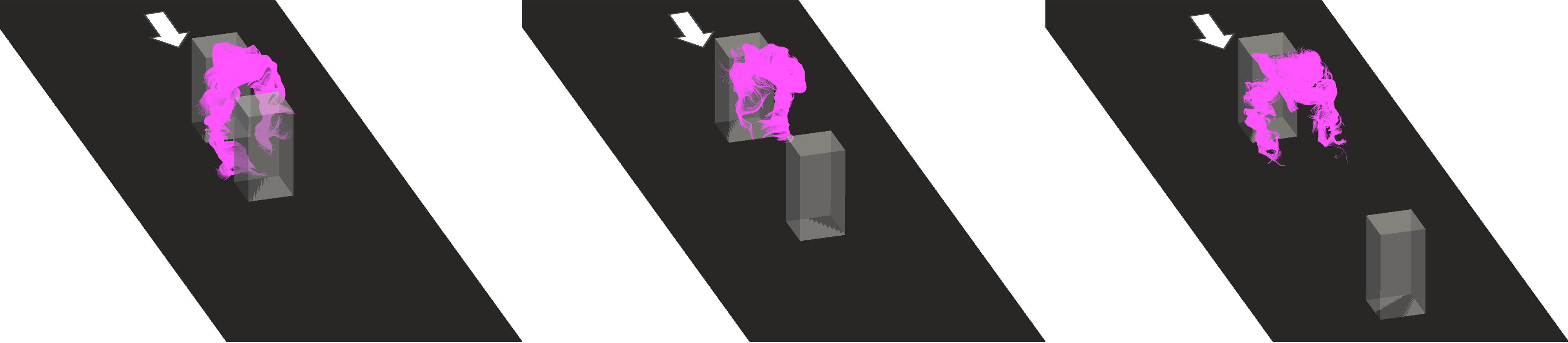}
         \caption{Vortex-generator mode}
     \end{subfigure}
     
     \begin{subfigure}[b]{\textwidth}
         \centering
         \includegraphics[width=\textwidth]{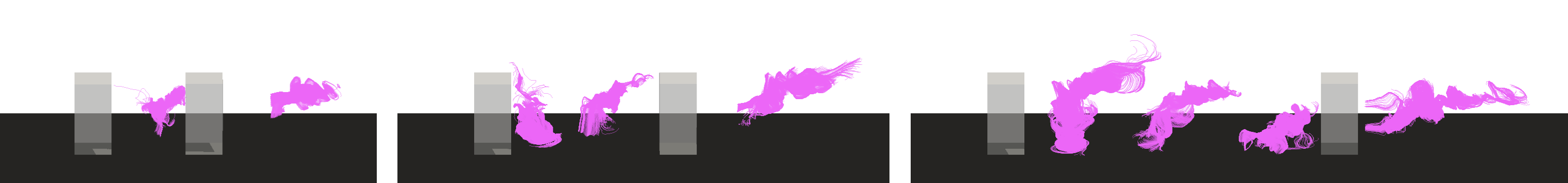}
         \caption{Vortex-breaker mode}
     \end{subfigure}
     \caption{Main flow patterns of the vortex-generator and vortex-breaker modes shown in Fig.~\ref{fig: DMD spectrum} visualized by means of streamlines for the three flow regimes. The arrows indicate the direction of the flow in each panel. Note the arch-shaped structure on the leeward side of the upstream obstacle and the helicoidal flow structures in between the obstacles.}
     \label{fig: HODMD streamlines}
\end{figure}

The wall-normal velocity component of mode A exhibits a three-dimensional pattern on the upper windward side of the upstream obstacle, which is shared among all the flow regimes. This cap-like structure created on top of the first building interacts primarily with the roof of the arch vortex. At this location, the flow experiences a high-velocity region due to the impact of the flow over the edge and the shear layer on the upper part of the obstacle. This region is then followed by another fluctuating part in the wall-normal direction in between the obstacles. In addition, a similar cap-like structure is observed in the IR case for the downstream obstacle. 
Finally, similar to the wall-normal component, the flow encounters a high-spanwise velocity gradient owing to the effect of the flow over the edges of the first obstacle, which tends to deviate the flow towards the outer regions of the domain. Furthermore, the second obstacle has comparable structures on its windward lateral edges as a result of the flow reattachment that occurs in the IR case~\cite{energies2021}. On the other hand, due to the slight interaction of the freestream flow coming from the lateral sides in the spanwise direction with this region, no fluctuating zones develop in between the obstacles. This characteristic distinguishes A modes from B modes, where the latter exhibit substantial spanwise variations between the obstacles. This may suggest that the structures of A modes could be connected with the formation of the vortical structures in this region. \alvaro{In particular, Fig.~\ref{fig: HODMD streamlines} shows the streamlines associated with this type of mode, where it can be noticed the high resemblance of this mode with the arch-vortex structure shown in Fig.~\ref{fig: Arch Vortex}}. It can then be stated that the flow features that predominate in the vortex-generator mode are \rev{resemblant of those of the mean flow}, emphasizing the idea of being a formation-type mode.

\alvaro{To conclude, the most unfavorable characteristics of the generation mechanisms with regards to the pollutant dispersion may appear to be the dome and the cap since they are linked to a delay in the pollutant vertical escape. Columns appearing on both sides of the buildings for the spanwise component may have less influence. They prevent the flow from spilling over the sides of buildings, but not from leaving the urban area. Further research is ongoing to confirm the relationship of the previous mechanisms with the dispersion of pollutants within urban environments. However, this topic remains out of the scope of this paper.}

\subsection{Breaking process of the main vortices}

The three-dimensional structures of the vortex-breaker mode are illustrated in Fig.~\ref{fig: HODMD main modes} (right). This mode is closely connected with the wake as opposed to what was observed in the previous case. In this type of mode, three additional structures can be distinguished for each velocity component: a turbulent wake, coherent cluster between the buildings, and arrowhead-like shapes. For all of the flow regimes investigated here, the spanwise-fluctuating areas are shown to occupy the whole region between the obstacles. As a result, increasing the distance between the obstacles leads to a greater number of these arrowhead-like structures in the spanwise direction: in the IR case, up to three alternating structures may be seen, whereas only one can be noticed in the SF regime. Conversely, the streamwise component of the present mode does not exhibit the same behavior; the coherent structures formed on the leeward side of the upstream obstacle remain unchanged for all three flow regimes. However, another oscillating zone arises connected to the downstream obstacle, the position of which is modified among the various regimes. Finally, in terms of velocity in the wall-normal direction, \alvaro{the high-velocity clusters appear within the region in between the buildings}. As the separation increases, the flow interacts inside the canopy with considerably more significance, resulting in larger flow structures for both the WI and IR cases. These structures are consistent with the results of Monnier et al.~\cite{Monnier2018} \rev{(see Fig.~\ref{fig: Experimental data})}, where strong streamwise fluctuations on both lateral sides and a high turbulent spanwise region near to the windward side of the downstream obstacle were reported. The arch vortex is known to exist between these regions, and these structures will be related to the process of breaking rather than creation, owing to its location on the wake. \alvaro{The interaction of the previous structures relates to the creation of a tunnel-shaped vortical structure between the buildings which might be responsible for the breaking process of the arch vortex. This flow mechanism is clearly elucidated by means of the streamlines patterns depicted in Fig.~\ref{fig: HODMD streamlines}. While the first mode resembles the arch-vortex structure, the second exhibits a helicoidal tunnel-shaped flow pattern in the region in between the obstacles, owing to the increased correlation in the spanwise direction. The location of these structures perfectly matches the gaps in between the velocity-fluctuating regions in the streamwise and spanwise directions. Therefore, since these \rev{velocity-fluctuating} regions define the location of such \rev{tunnel-shaped flow} patterns, the number of structures is modified from case to case: up to three structures are observed in between the buildings for the IR regime. Consequently, the interaction of the structures of B modes within this region results in a mixing procedure, leading to the breaking process of the vortical structures that are generated by A modes.}

\subsection{Interaction between vortex-generator and vortex-breaker modes\label{sec: Interaction}}

\begin{figure}[ht!]
     \centering
     \begin{subfigure}[b]{\textwidth}
         \centering
         \adjincludegraphics[width=0.95\textwidth,trim={{.05\width} {.025\width} {.05\width} {.02\width}}]{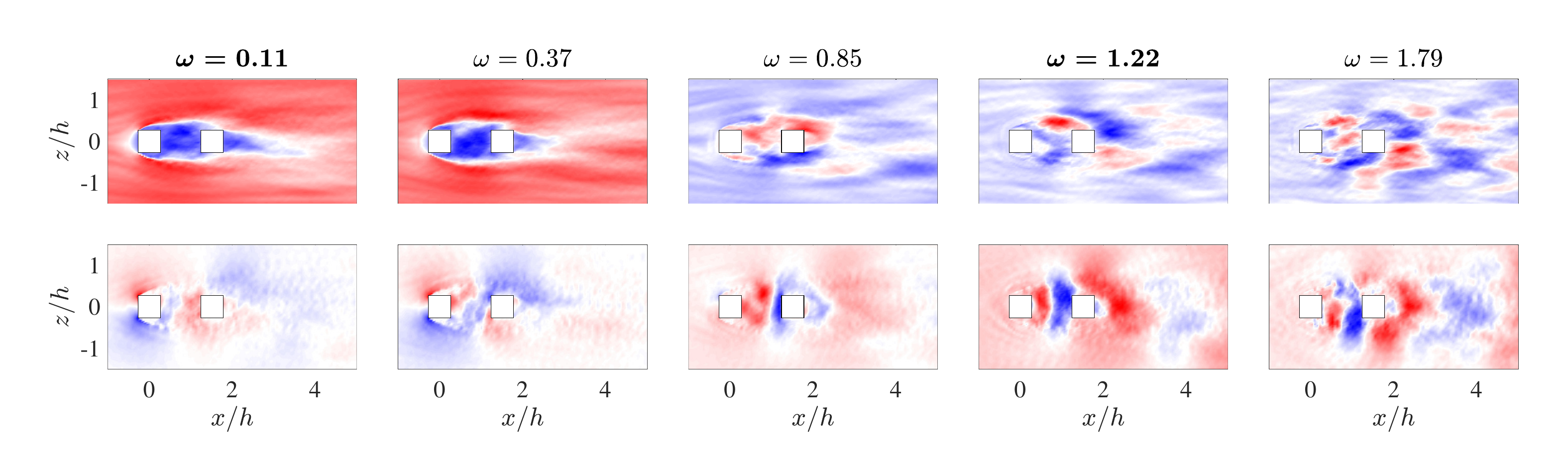}
         \caption{Skimming flow\vspace{0.25em}}
     \end{subfigure}
     \begin{subfigure}[b]{\textwidth}
         \centering
         \adjincludegraphics[width=0.95\textwidth,trim={{.05\width} {.025\width} {.05\width} {.02\width}}]{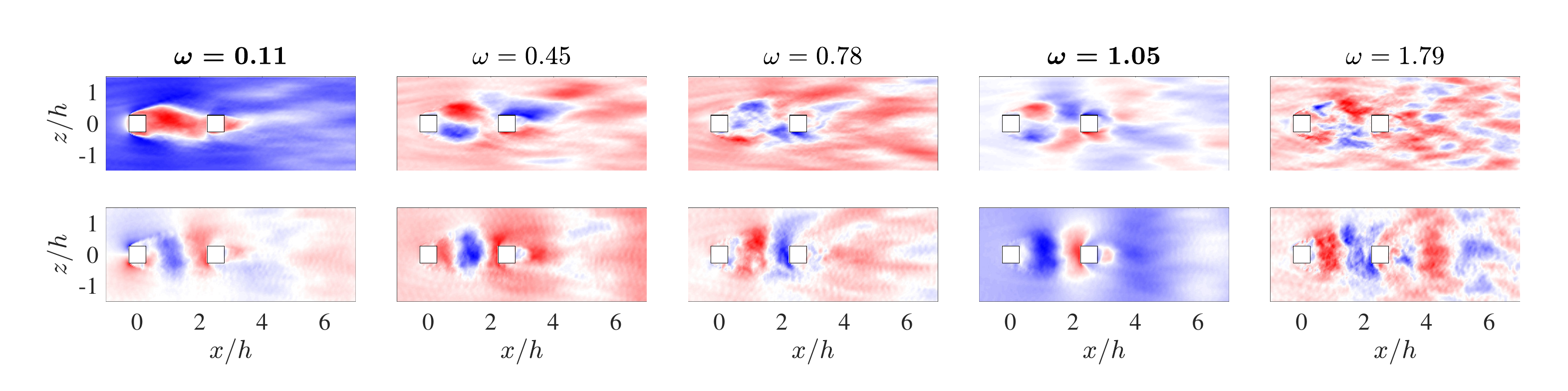}
         \caption{Wake interference\vspace{0.25em}}
     \end{subfigure}
     \begin{subfigure}[b]{\textwidth}
         \centering
         \adjincludegraphics[width=0.95\textwidth,trim={{.05\width} {.025\width} {.05\width} {.02\width}}]{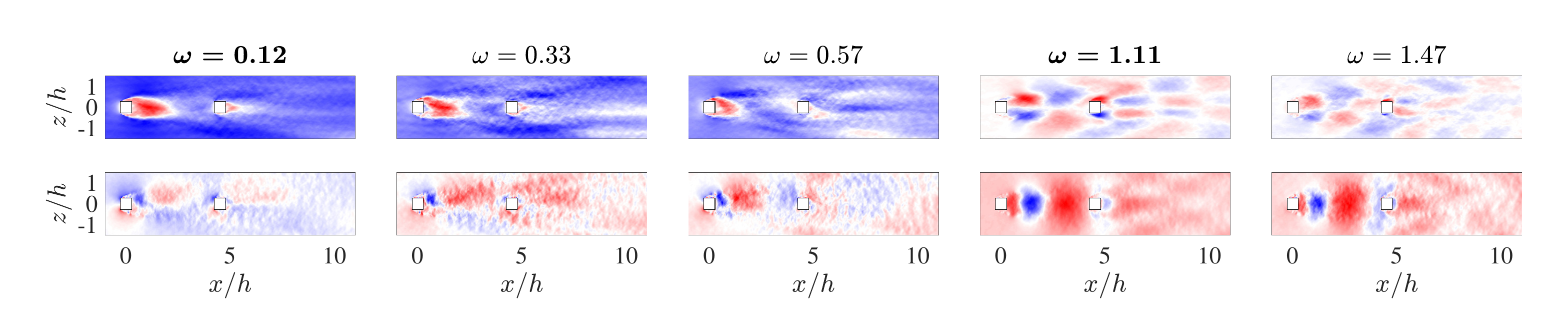}
         \caption{Isolated roughness}
     \end{subfigure}
     \caption{\rev{HO}DMD modes of the streamwise and spanwise velocity fields with selected frequencies, represented at $y/h=0.25$ for the different flow regimes. Contours of the velocity of the modes are normalized with the $L_\infty$-norm and vary between $-1$ (blue) and $+1$ (red). The bold-face frequencies represent the vortex-generator and the vortex-breaker modes, modes A and B, respectively. The rest of the modes are the result of the interaction between the above-described modes and are known as harmonic modes.}
     \label{fig: HODMD harmonic modes}
\end{figure}

Finally, the three-dimensional structures of the most significant \rev{HO}DMD modes can be compared to those of the various modes identified by the algorithm. Specifically, the following lines will be dedicated to the classification of the modes in vortex-generator or breaking-vortex modes based on resemblance with the prior patterns. Note that this classification is driven by the MAC analysis stated in previous sections, which allows for selecting the more robust modes identified by HODMD. Fig.~\ref{fig: HODMD harmonic modes} shows a contour representation at $y/h=0.25$ of the \rev{HO}DMD modes presented in Fig.~\ref{fig: DMD spectrum}. The main A and B modes are highlighted in bold and their two-dimensional structures can be compared with the previously-discussed three-dimensional ones. From these structures, a limit frequency can be established such that greater-frequency values result in \alvaro{flow structures which are more related to the vortex-breaking process}. For these modes, the MAC analysis gives values which are closer to 1 when compared to B modes, \rev{indicating similarity in shape with B modes}.

For the SF case, the mode ${{\omega_m}} = 0.37$ still exhibits some flow trapped in between the obstacles and high spanwise fluctuations on the lateral edges of the first obstacle. Even though the flow in this region appears to be modified by the slight interaction with the surrounding flow, this mode can be thought of as a vortex-production mode with a different production mechanism. Higher-frequency modes (${\omega_m}>0.85$) are characterized as breaker modes since both the streamwise and spanwise components share the same flow features as the main B mode (with frequency $\omega_{m}=1.22$). Note as well that higher-frequency modes exhibit smaller spatio-temporal scales. This highlights the association of low-frequency modes with large flow scales (dominant patterns) and high-frequency modes with smaller turbulent structures. A similar conclusion can be extracted for the IR regime, where the threshold value is set for the mode with frequency ${\omega_m}=0.57$. This mode exhibits some flow structures around the upstream obstacle combined with particular features on the wake, which makes it a transitory mode between the vortex-generator and vortex-breaker modes. Finally, regarding the WI case, apart from the vortex-generator mode (${\omega_m}=0.13$), the lowest-frequency mode (${\omega_m}=0.45$) exhibits a flow pattern similar to that of the vortex-breaker mode. Therefore, in this situation, the threshold value should be set lower than this frequency, resulting in all modes fulfilling ${\omega_m}>0.45$ becoming of breaking-type. Note that for this case, modes with associated frequencies $\omega_m = 0.11$, $0.22$ and $0.33$, which are harmonics of the lowest-frequency mode, are examples of vortex-generator modes with a MAC value close to 1 when compared to the main A mode, \rev{indicating significant similarity with this mode}.

Knowledge of the mechanisms of generation and destruction of relevant vortical structures within urban flows provides sufficient information to be able to perform studies of pollutant dispersion within urban environments, so that ground-level concentrations significantly higher than those occurring in the absence of the building can be avoided. In this sense, for pollutants emitted at street level, the vortex-breaker mode could provide, at a high frequency, a higher interaction with the surrounding clean air, while for the arch-generating mode, the flow between buildings could be hardly influenced by the flow outside. Therefore, B-type modes could be connected to the promotion of the pollutant dispersion within cities, whereas the generation of A-type modes should be minimized owing to their low interaction with the surrounding atmosphere. Another important aspect closely related with the pollutant-dispersion aspect is the direction of the fluctuations. Regarding vortex-breaker modes, the tunnel-shaped structures \alvaro{shown in Fig.~\ref{fig: HODMD streamlines}}, mainly influenced by the arrow-shaped spanwise fluctuations, would disperse rapidly those pollutant emitted at the street level towards the atmosphere. However, the streamwise fluctuations, which increase close to the building walls, produce an increased concentration of pollutants within the city.\\

\subsection{Streamwise and spanwise-periodic structures}

\begin{figure}
	\centering
	\begin{subfigure}[b]{0.49\textwidth}
		\includegraphics[width=\textwidth]{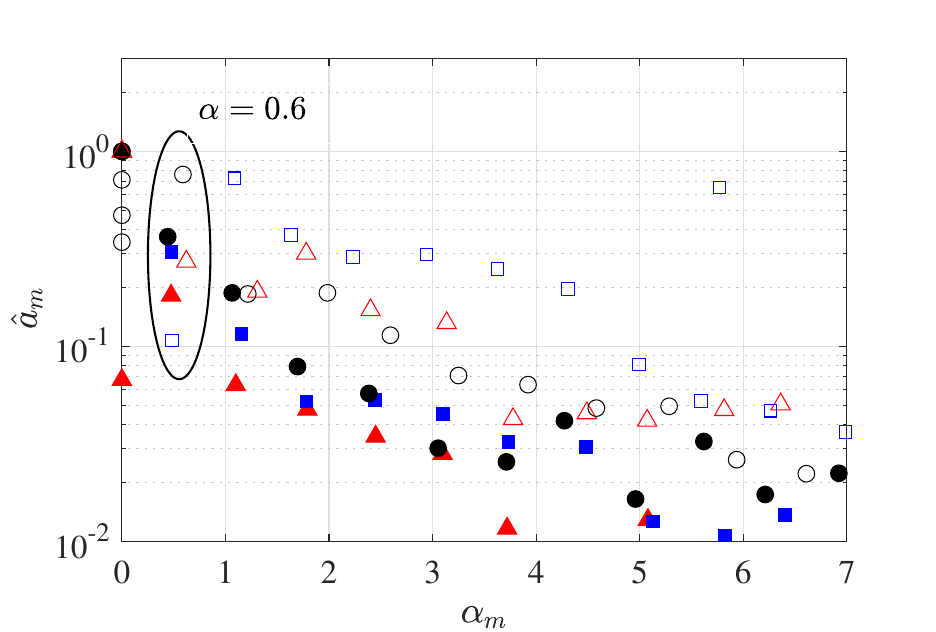}
	\end{subfigure}
\hfill
	\begin{subfigure}[b]{0.49\textwidth}
		\includegraphics[width=\textwidth]{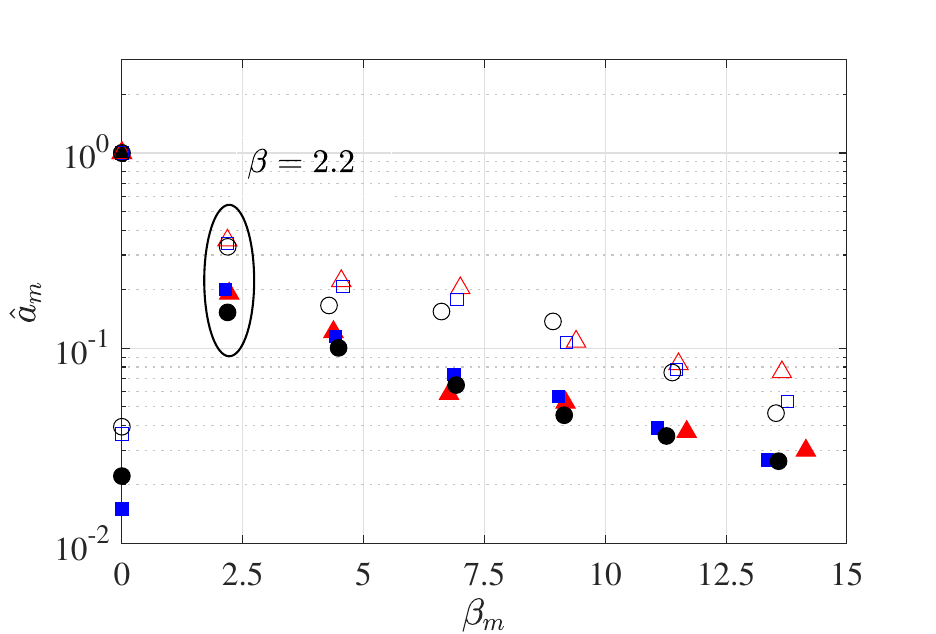}
	\end{subfigure}
	\caption{Spectra of the STKD modes for the (left) streamwise and (right) spanwise directions. Amplitude scaled with the maximum value ($\hat{a}_m=a_m/a_0$) versus wavenumber, ${\alpha}_m$ for the $x$-analysis and ${\beta}_m$ for the $z$-analysis, computed for ({red}) skimming flow, ({blue}) wake interference and ({black}) isolated roughness. The filled points represent the spatial modes obtained from the generator temporal modes while the empty markers represent the ones obtained from the breakers. The modes represented here are the result of a calibration process of the user parameters, from where $\varepsilon_\text{SVD} = \varepsilon_\text{DMD} = 10^{-4}$, $d=5$ for every case.}
	\label{fig: STKD spectrum}
\end{figure}

After discussing the different \rev{possible} generation and breaking mechanisms of the main vortical structures associated with the temporal modes obtained by HODMD, the STKD algorithm is applied to these modes in order to obtain the spatio-temporal modes, with the aim of understanding \rev{more in detail} the mentioned mechanisms and how they link with the physics of the problem.

As already shown in \S\,\ref{sec: STKD methodology}, \rev{the modal} expansion can be applied to different spatial directions and, in this work, we analyze both the streamwise and spanwise components. Once the STKD is applied to the temporal HODMD modes, we obtain a spectrum of the spatio-temporal modes as can be seen in the Fig.~\ref{fig: STKD spectrum} for both directions, where the modes obtained for the $x$-direction are called X modes and the ones obtained in the $z$-direction are the Z modes. The dominant wavenumbers are $\alpha_m=0.6$ for the X modes, and $\beta_m=2.2$ for the Z modes. In general, it can be observed that the spatial modes obtained from the A modes have lower amplitude than the ones obtained from the B modes. After studying the results obtained from the spatio-temporal modes, it can be observed that the X modes are connected to the mechanisms of the breaking process and the Z modes show the \rev{resulting structures from these breaking mechanisms}. 

\begin{figure}[t]
	\centering
	\includegraphics[width=\textwidth]{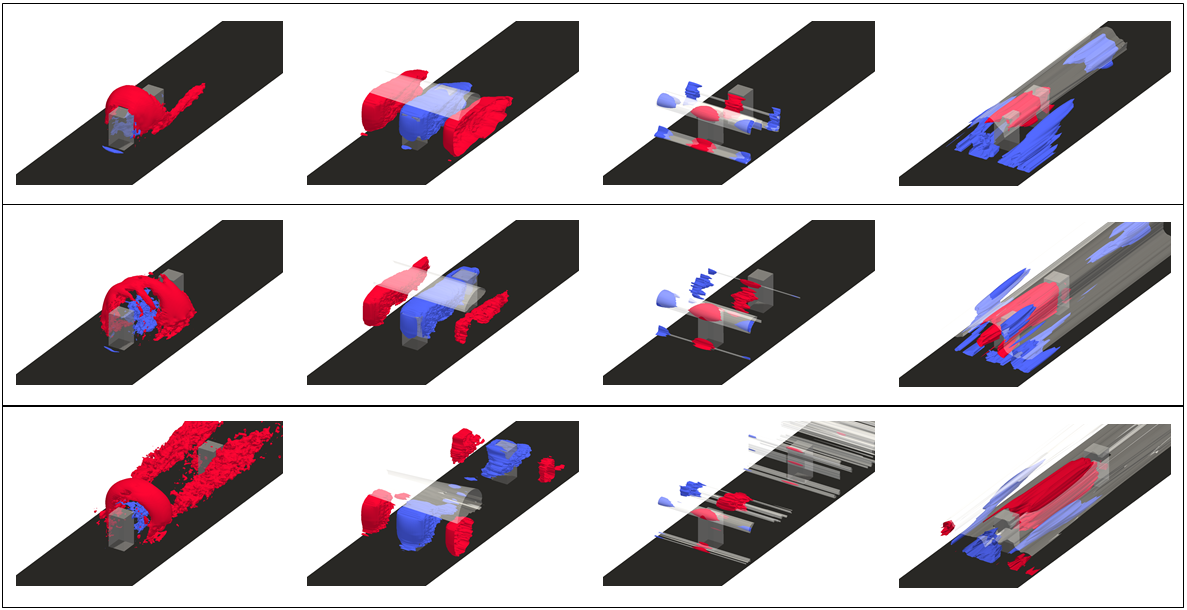}
	\caption{Three-dimensional iso-surfaces of the different spatio-temporal modes obtained from the generator modes. (Top), (middle) and (bottom) show the SF, WI and IR cases respectively, while each column represents a different mode: the first column shows the streamwise velocity of the temporal generator mode, the second and third columns display the streamwise and wall-normal velocity of the dominant Z mode with a wavenumber of $\beta_m=2.2$, and the last column represents the streamwise velocity of the dominant X mode with a wavenumber of $\alpha_m=0.6$. Red and blue denote positive and negative velocities respectively, and the white-transparent structures represent the mean-flow of the spatio-temporal modes with zero wavenumber.}
	\label{fig: STKD generators}
\end{figure}  

Fig.~\ref{fig: STKD generators} depicts the different structures appearing in the STKD analysis when applied to generator modes. The interaction zone for all the displayed modes is near and between the buildings, hence, the shedding of the arch vortex still has not started. \rev{Regarding the unsteady modes, it is possible to identify different types of flow patterns:} for the Z modes, the \rev{patterns} that appear in the streamwise component of the velocity surround both buildings for the SF and WI regimes while these \rev{patterns} are divided \rev{in the IR case}, creating a new structure in the second building. This phenomenon occurs \rev{due to the wave-like topology of the flow in} the IR regime, \rev{where the free stream has enough distance between the buildings to adapt}. Something similar happens for the wall-normal component \rev{in the IR case}, where the cap-like structure starts to appear again in the second building. Also, a phase shift appears for the \rev{patterns} formed in the streamwise component of the Z mode between the real and imaginary part (the imaginary part is not shown for the sake of brevity), \rev{showing} that these structures are traveling waves along the spanwise direction. \rev{The same applies to the X modes, where the phase shift between the real and imaginary part shows that these structures are traveling waves along the streamwise direction. Regarding the interaction between the mean flow and the unsteady modes, for the streamwise component of the Z modes, the zero-frequency mode or mode 0 appears between the buildings interacting with the structures of the dominant mode, suggesting its connection with the presence of the arch vortex. Similar phenomena are observed for the wall-normal Z modes, although the interaction occurs on the top and bottom of the first building, suggesting that there is connection between the cap recirculation structure and the horseshoe vortex. Finally, the streamwise component of the X modes shows that the mode 0 surrounds the two buildings interacting with the unsteady modes, suggesting its relationship with the arch vortex again.}

\begin{figure}[t]
	\centering
	\includegraphics[width=\textwidth]{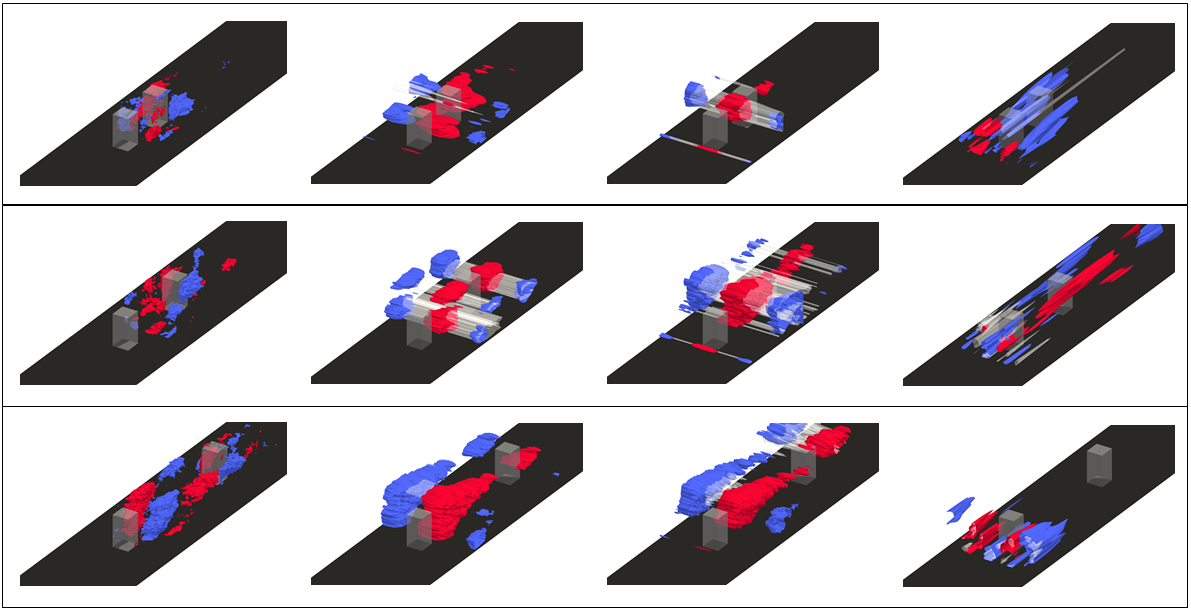}
	\caption{Three-dimensional iso-surfaces of the different spatio-temporal modes obtained from the breaker modes. (Top), (middle) and (bottom) show the SF, WI and IR cases respectively, while each column represents a different mode: the first column shows the streamwise velocity of the temporal breaker mode, the second and third columns display the streamwise and wall-normal velocity of the dominant Z mode with a wavenumber of $\beta_m=2.2$, and the last column represents the streamwise velocity of the dominant X mode with a wavenumber of $\alpha_m=0.6$. Red and blue denote positive and negative velocities respectively, and the white-transparent structures represent the mean-flow of the spatio-temporal modes with zero wavenumber.}
	\label{fig: STKD breakers}
\end{figure}  

Regarding Fig.~\ref{fig: STKD breakers}, where vortex-breaker modes are shown, the structures of the Z modes continue to appear between the buildings rather than on the sides of the buildings, as expected. The reason behind this phenomenon lies in the strong streamwise influence present in the flow, making the breaking mechanisms appear near the arch vortex. The X modes display large streamwise structures on the sides of the buildings (except in the IR case), \rev{suggesting the connection with} the destruction of the main vortices and \rev{generation of} the turbulent wake. \rev{With regards to} the Z modes, a large \rev{structure} appears between the buildings after the arch vortex, suggesting \rev{a possible connection} with the shedding of the arch vortex. \rev{Regarding the zero-frequency mode, it begins to appear between the buildings with low intensity for the SF, it enlarges and gains relevance in the flow for the WI case, while for the IR case its presence decreases. However, the dominant mode becomes important, especially in the IR, suggesting its connection with the wake that forms downstream of the buildings. For the vortex-breaker case, the interaction between the zero-frequency mode and the dominant frequency mode is present, however, as the intensity of the structures related to the unsteady modes is considerably higher than in the vortex-generator modes, the interaction has a lower relevance.}

\begin{figure}
  \centerline{\includegraphics[width=\textwidth]{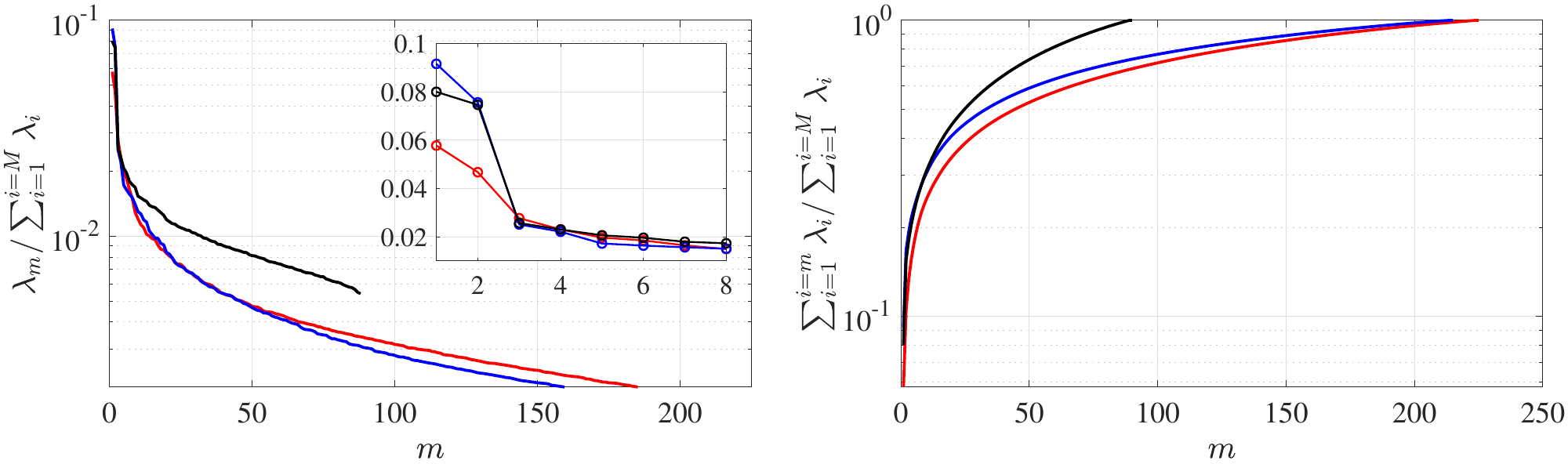}}
  \caption{\alvaro{Eigenvalues $\lambda_m$ (left) and cumulative sum of the eigenvalues $\sum_{i=1}^{i=m} \lambda_i$ (right) spectrum normalised with the total energy of the eigenvalues $\sum_{i=1}^M \lambda_i$} of the POD modes corresponding to the complete set of velocity components (streamwise, wall-normal and spanwise) of the ({red}) skimming-flow, ({blue}) wake-interference and ({black}) isolated-roughness regimes. The number of modes is denoted with $m$ and the total coincides with the number of columns of the snapshot matrix, i.e. $225$, $215$ and $90$, respectively.}
\label{fig: Singular Values}
\end{figure}

\section{Energy analysis} \label{sec: POD}

In this section, we analyze the same urban database discussed in \S\,\ref{sec: Numerical simulations} using POD. \alvaro{In Fig.~\ref{fig: Singular Values}, we show the eigenvalues $\lambda_m$ (left) and cumulative sum of the eigenvalues $\sum_{i=1}^{i=m} \lambda_i$ (right) spectrum normalized with the total energy of the eigenvalues $\sum_{i=1}^M \lambda_i$ of the POD modes corresponding to the complete set of velocity components of the three reference regimes.} The objective is to identify energy gaps that make some modes energetically more relevant than others. This energy gap is noticed between the second and third modes for all regimes, which highlights that the first two modes contain the \rev{most} relevant information of the flow. \alvaro{A second but smaller energy gap is appreciated between the fourth and fifth modes. These constitute the set of POD modes that will be compared with the main HODMD modes in the present section. A separated analysis conducted on the three components of the velocity revealed that the wall-normal component only accounts for $15\%$ of the total energy content of the flow}. This means that the influence of the wall-normal velocity fluctuations is lower than that of the other velocity components. This is consistent with the findings of Monnier et al.~\cite{Monnier2018} \rev{depicted in Fig.~\ref{fig: Experimental data}}, who discovered that for a zero-incidence angle, the streamwise and spanwise fluctuating components are more significant than the wall-normal component. On the basis of the above, only the streamwise and spanwise components will be further studied for the three flow regimes.

\begin{figure}[t]
     \centering
     \begin{subfigure}[b]{\textwidth}
         \centering
         \includegraphics[width=\textwidth]{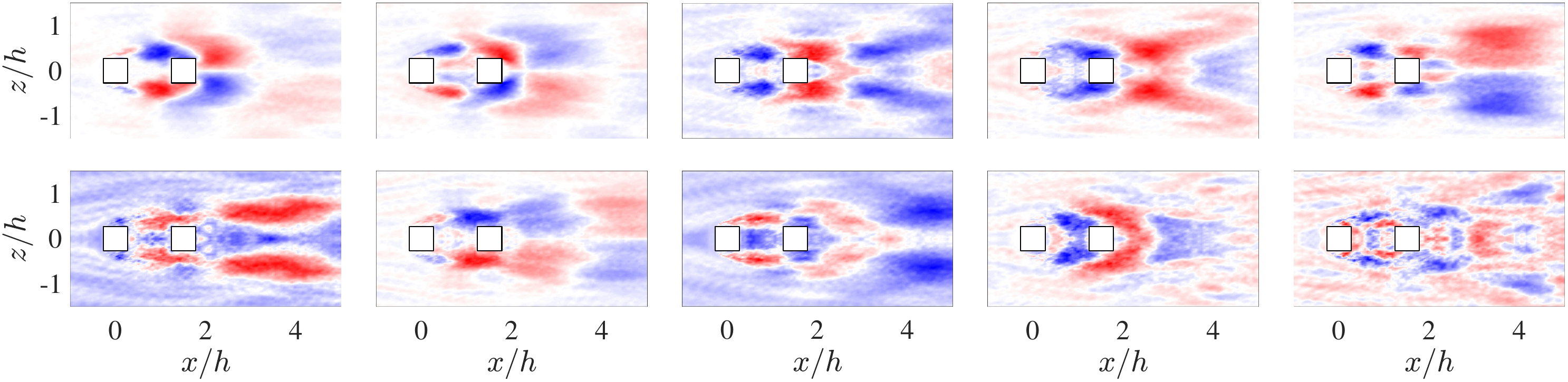}
         \caption{Skimming flow\vspace{0.75em}}
     \end{subfigure}
     \begin{subfigure}[b]{\textwidth}
         \centering
         \includegraphics[width=\textwidth]{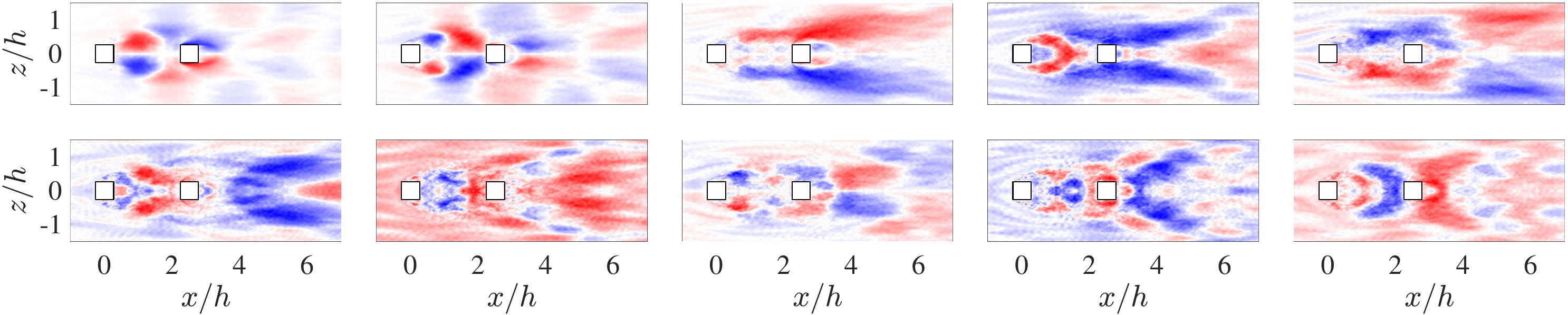}
         \caption{Wake interference\vspace{0.75em}}
     \end{subfigure}
     \begin{subfigure}[b]{\textwidth}
         \centering
         \includegraphics[width=\textwidth]{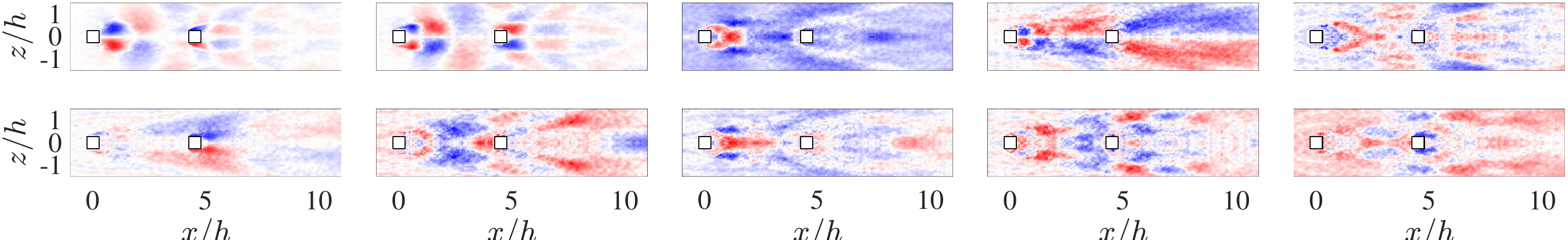}
         \caption{Isolated roughness}
     \end{subfigure}
     \caption{POD orthogonal basis of the streamwise velocity fields at $y/h=0.25$ for the different flow regimes. \alvaro{For each regime, from the upper left to the lower right, first to ten modes are sequentially presented}. Contours of the velocity of the modes are normalized with the $L_\infty$-norm and vary between $-1$ (blue) and $+1$ (red).}
     \label{fig: POD modes streamwise}
\end{figure}

\begin{figure}[t]
     \centering
     \begin{subfigure}[b]{\textwidth}
         \centering
         \includegraphics[width=\textwidth]{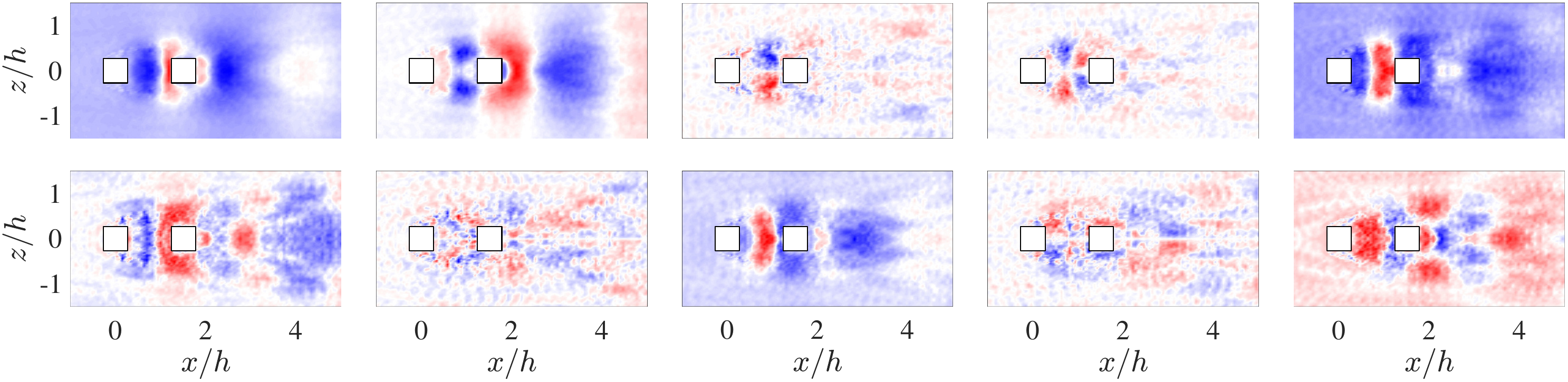}
         \caption{Skimming flow\vspace{0.75em}}
     \end{subfigure}
     \begin{subfigure}[b]{\textwidth}
         \centering
         \includegraphics[width=\textwidth]{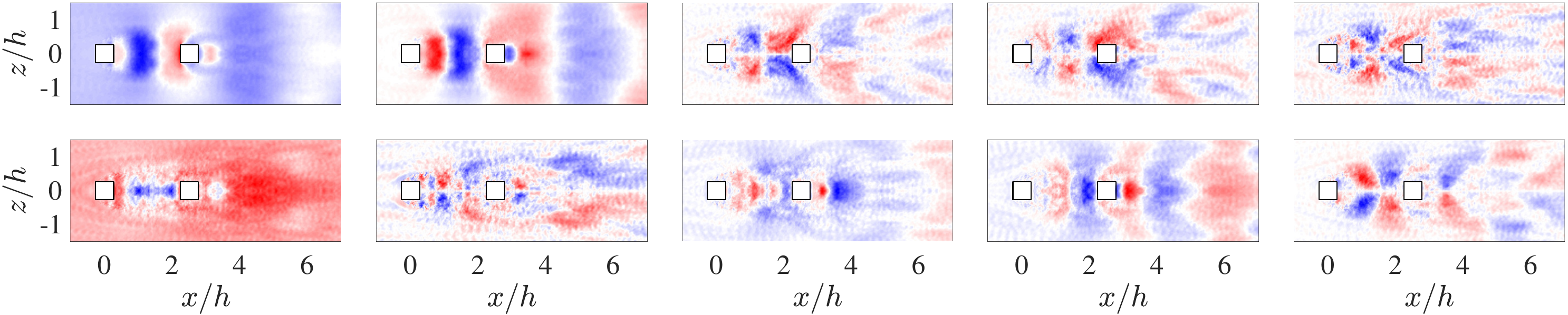}
         \caption{Wake interference\vspace{0.75em}}
     \end{subfigure}
     \begin{subfigure}[b]{\textwidth}
         \centering
         \includegraphics[width=\textwidth]{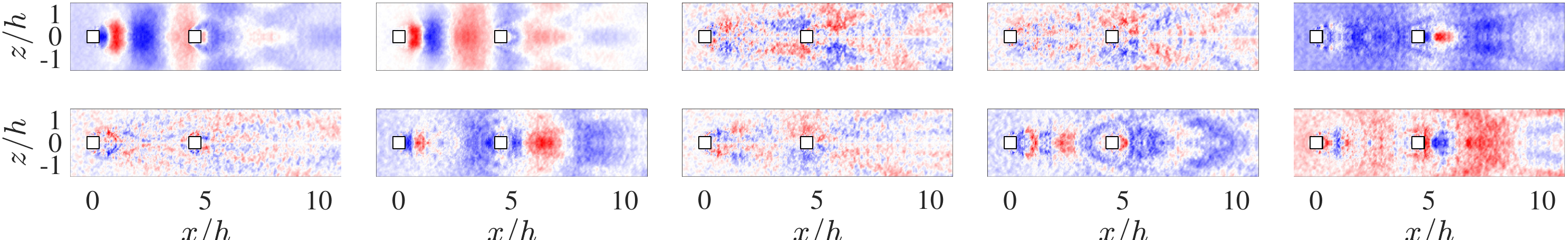}
         \caption{Isolated roughness}
     \end{subfigure}
     \caption{POD orthogonal basis of the spanwise velocity fields at $y/h=0.25$ for the different flow regimes. \alvaro{For each regime, from the upper left to the lower right, first to ten modes are sequentially presented}. Contours of the velocity of the modes are normalized with the $L_\infty$-norm and vary between $-1$ (blue) and $+1$ (red).}
     \label{fig: POD modes spanwise}
\end{figure}

\begin{figure}
    \centering
    \includegraphics[width=\textwidth]{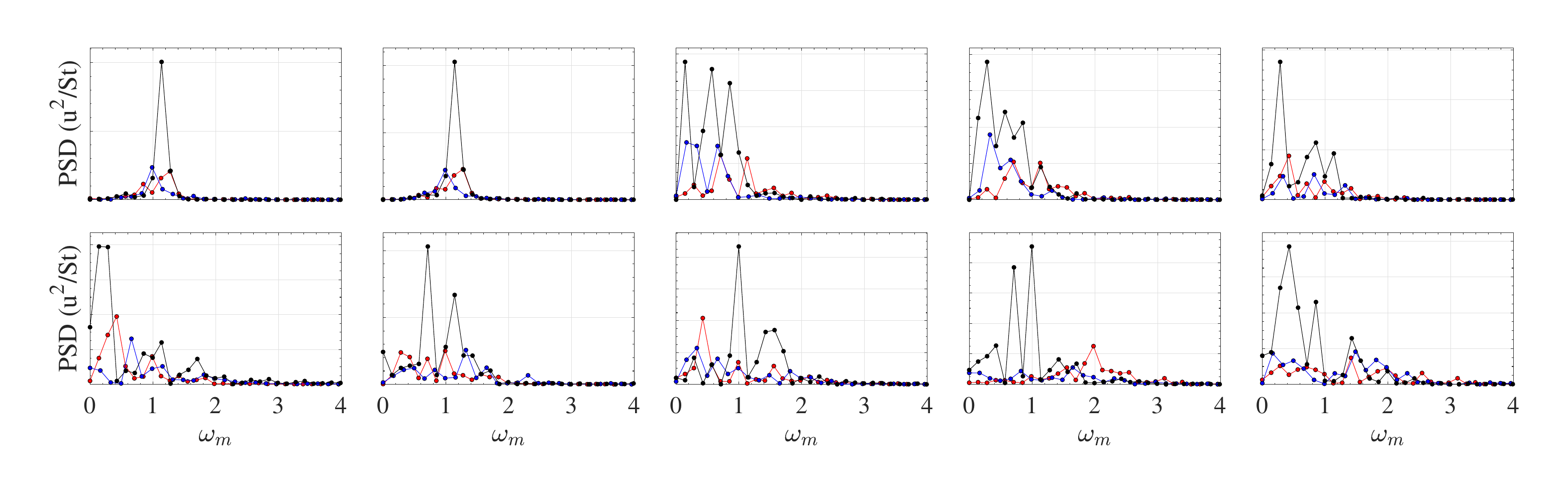}
    \caption{Power-spectral density scaled with the Strouhal number $St=f h/U_\infty$ of the temporal coefficients associated to POD modes, where $f$ is the characteristic frequency of each mode: ({red}) skimming-flow, ({blue}) wake-interference and ({black}) isolated-roughness regimes. As in Figs.~\ref{fig: POD modes streamwise} and \ref{fig: POD modes spanwise}, from upper left to the lower right, first to ten modes are sequentially depicted. \alvaro{PSD is calculated using $N=128$ samples for SF and WI regimes and $N=64$ samples for IR regime, with a window overlap of $50\%$ for every case.}}
    \label{fig: FFT POD All}
\end{figure}

Figs.~\ref{fig: POD modes streamwise} \rev{and \ref{fig: POD modes spanwise}} depict the orthogonal POD basis for the first \rev{ten} modes of the streamwise and spanwise velocity fields corresponding to the three flow regimes. \alvaro{In addition, an analysis of the temporal coefficients associated with these modes is performed in the frequency domain through the fast-Fourier-transform (FFT) method~\cite{cooley1965}. Fig.~\ref{fig: FFT POD All} depicts the power spectrum for the first five POD modes in all the flow regimes. We can classify the time coefficients linked to each spatial mode into low- and high-frequency phenomena in the frequency range $\omega_m = \left[0,2\right]$ using this modal-decomposition technique, whose characteristics are very relevant to the vortex-generating and -breaking processes, respectively. In fact, the first two modes, with associated frequencies that are similar to those of the HODMD B modes ($\omega_m \approx 1$), are characterized by high-velocity streamwise fluctuations on both sides of the buildings; these are complemented by spanwise-velocity fluctuations in the area between the obstacles for each flow regime.} These regions match with the high-turbulent-kinetic-energy (TKE) regions of the streamwise component identified by Monnier et al.~\cite{Monnier2018} for an array of building-like blocks, \rev{see Fig.~\ref{fig: Experimental data}}. However, the main differences among regimes depend on the position of the secondary structures, which are associated with the downstream block. For instance, while these streamwise fluctuating regions span the zone in between the obstacles for the skimming flow and the wake interference cases, in the isolated roughness, they are only located on the immediate leeward side of the upstream block. In such a fashireon, increasing the separation of the obstacles does not yield more streamwise fluctuating regions, at least for the more energetic modes, as it occurs in a vortex-shedding case. \rev{Furthermore, the spatial structures of the first two modes are observed to be the same for the three flow regimes, except for a shift in phase: they are both antisymmetric about the $z$-axis and they both represent fluctuating regions on the wake with matching frequency values. This is an evidence that these modes represent a wave-like periodic structure of the flow that develops in the streamwise direction.}

On the other side, increasing the distance between the obstacles increases the number of high-intensity flow structures of spanwise fluctuating regions, although the width of these structures decreases. These results could be connected to the interplay of lateral flow within the canopy, which suggests that the arch vortex formed on the leeward side of the upstream obstacle is shattered. \alvaro{A tunnel-shaped structure is then produced as a result of the interaction between these two different types of structures, a fact that suggests that it is this process that breaks the main flow structures (see Fig.~\ref{fig: HODMD streamlines}). Therefore, the vortex-breaking process has been identified as the most energetically relevant mode present in the flow field.}

The third and fourth POD modes are related to the vortex-generating modes, or the A modes, because of their low-frequency behavior ($\omega_m < 0.8$). In this case, the streamwise component demonstrates how a dome-like structure encloses the area between the obstacles and expands further in the wake, with low spanwise fluctuations. Due to the resemblance with the time-averaged field, a generating process for these structures is implied. \rev{This is consistent among the three flow regimes; however, for the IR regime, these structures appear closer to the first obstacle and a clear interaction with the second obstacle is not observed. A similar conclusion was drawn for the most energetic modes, a fact that highlights the independence of flow around both obstacles in the IR regime.} 

The fifth mode, which produces flow structures as a result of the interaction of the aforementioned low- and high-frequency modes, can be thought of as a transitional mode between them. Therefore, higher-order modes exhibit flow structures which might result from the combination of the previous modes and if even higher-order modes $\left(m > 10\right)$ were studied, certain high-frequency phenomena would be captured due to the smaller turbulent-flow scales associated with them.  \rev{As a result, due to the wide range of frequencies in their spectrum, a clear comparison with HODMD modes is more difficult to be set for these higher-order modes since the former are associated with a single frequency value. Nevertheless, this transitional behavior is also observed in the spatial structures from the fifth to tenth modes in Figs.~\ref{fig: FFT POD All} and \ref{fig: POD modes spanwise}, where the wide range of structures found might be regarded as a combination of vortex-breaker and vortex-generator features. Remarkably, for some of the higher-order modes of the isolated roughness regime, e.g. modes 6 and 10, individual structures are observed around the downstream obstacle. Again, this provides more evidence of the independence of the flow in this flow regime, but also of the ability of the method to discern between flow patterns around both obstacles, a characteristic that was not found for the HODMD modes.}

\section{Summary and conclusions} \label{sec: Conclusions} 

A simplified urban environment model consisting of an array of two buildings with variable spacing ratios was examined using high-fidelity simulations. These simulations were carried out to provide a complete physical description of the fundamental mechanisms controlling the dynamics in different urban streets. The aim was to provide an insightful analysis of the physics of the flow within environments that approximately reproduce some aspects of different types of urban areas. The growing expansion of cities boosts the search for physical models capable of reproducing the pollutant and thermal distributions within cities, although we consider simplified versions of those flow cases. Here, the three-dimensional flow patterns responsible for pollutant dispersion have been characterized. Isosurfaces and contour slices were used to demonstrate the complicated flow behavior of the modes identified by the POD and HODMD algorithms.
The results show that the flow behavior can be split into low- and high-frequency phenomena, each with significant consequences related to the formation and destruction of vortical structures such as the arch or horseshoe vortices. Low-frequency modes are named vortex-generator (A) modes since their associated structures have been related to the mechanism triggering the formation of the arch vortex formed on the leeward side of both buildings. These structures are particularly noticeable on the windward side of the upstream obstacle for the wall-normal and spanwise velocity components and the leeward side for the downstream one, which defines the location and shape of the arch vortex. Furthermore, this location is kept constant among the different flow regimes, which highlights the idea that the process of formation of the arch vortex does not strongly depend on the separation between the obstacles.

In addition, HODMD identifies a high-frequency mode that correlates with the largest-amplitude mode in all cases. Because of the streamline flow patterns in the intermediate section of the obstacles, they are referred to as vortex-breaker modes. The large amplitude of these modes emphasizes their importance in this type of simplified urban environment. Indeed, their structures are linked to the first two highest-energy POD modes. Furthermore, as the separation increases, the flow becomes more correlated in the spanwise direction, owing to the more significant interaction of the flow with the wake layer inside the canopy. This effect yields to fluctuating velocity regions occupying the whole section between the obstacles, thus being associated with the destruction of the vortical structures rather than their formation.
Besides, the vortex-breaker modes will be responsible for the emergence of high-TKE regions on both sides of the obstacles and on the windward side of the downstream obstacle. Interestingly, these results are consistent with the wind-tunnel results of Monnier et al.~\cite{Monnier2018}, performed on a more complex urban environment. Therefore, the conclusions of the present work could potentially be relevant to more realistic urban environments by considering that the turbulence levels from one street to the next one are expected to decrease significantly. It is also interesting to note that the results provided by both POD and HODMD show that the wall-normal velocity component does not significantly influence the more prominent structures as the streamwise and the spanwise components do.

Regarding the results obtained from the STKD analysis, the Z modes display the results of the mechanisms obtained in the X modes about the generation and destruction of the coherent structures. When analyzing the temporal generator modes, the X and Z modes show a phase shift between the real and imaginary parts. Consequently, traveling waves appear in each direction. In addition, the symmetry is conserved, and the influence area shows that the structures are still unbroken. On the other hand, when the STKD analysis is applied to the temporal breaker modes, the X modes show that the main structures are broken, and the structures causing this destruction are shaped as large streaks in the streamwise direction. Meanwhile, the Z modes show the \rev{resulting structures from these breaking mechanisms}.

From an environmental point of view, urban areas with highly-separated buildings, i.e. the isolated-roughness regime, would exhibit much more interaction with clean air sources, thus enabling the rapid propagation of those pollutants emitted at the street level. However, power plants, commonly located close to urban centers, are also responsible for pollution issues within cities. In those cases, owing to the low interaction of the flow above the urban canopy with the streets, it would be convenient to decrease the separation between buildings, i.e. establishing the skimming-flow regime. In such a case, the air at the street level would also be in contact with clean sources of air through the arch-vortex legs.

\rev{We conclude our work with a brief discussion over the main limitations of the methods employed here. Firstly, a numerical data set of more than 200 three-dimensional snapshots for the three components of the velocity has been found to be accurate enough to represent the main large-scale structures of the flow. However, this number of fields has a direct impact on the required computational resources, especially given the amount of information employed for each field. Although some additional analyses have been performed with a slightly larger number of snapshots, these results remain to be confirmed with data sets encompassing a significantly larger number of snapshots. Additionally, higher-order modes could also be analysed: their associated structures might differ from those of the modes presented here, thus yielding to new types of modes. Other modal-decomposition techniques such as SPOD may also be employed with the objective of confirming the similarity of the obtained modes with the present spatio-temporal structures.}

\appendix

\section{Calibration process of the HODMD algorithm}
\label{Appendix: Calibration}

\begin{figure}[t]
    \centering
    \adjincludegraphics[width=0.825\textwidth,trim={{.05\textwidth} {.035\textwidth} {.05\textwidth} {.05\textwidth}}]{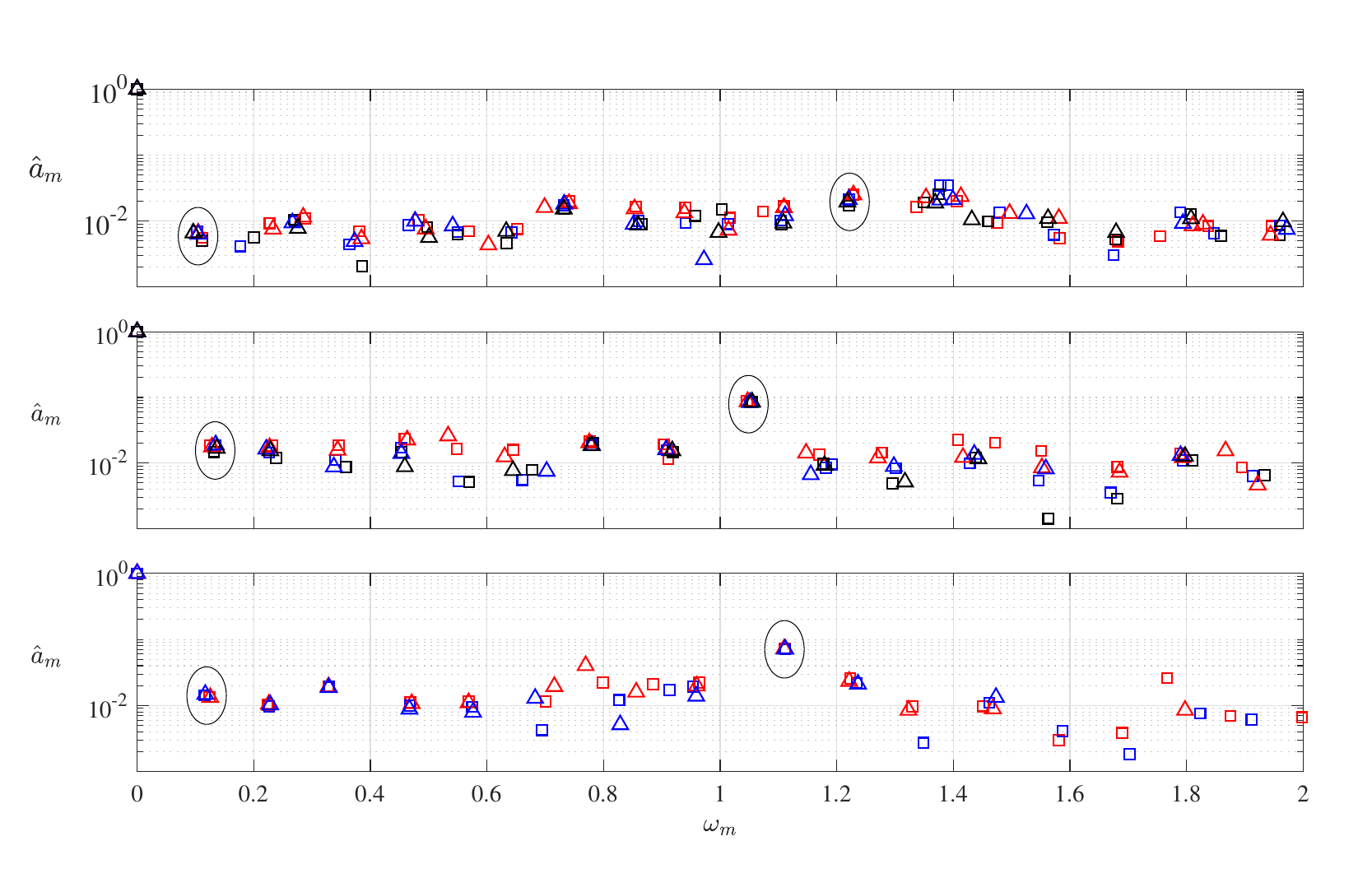}
    \caption[DMD-d calibration. Amplitude scaled with its maximum value ($\hat{a}_m=a_m/a_0$) versus frequency ${\omega}_m$ computed with different tolerances for ({top}) skimming flow, ({middle}) wake interference and ({bottom}) isolated roughness.]{DMD-d calibration. Amplitude scaled with its maximum value ($\hat{a}_m=a_m/a_0$) versus frequency ${\omega}_m$ computed with different tolerances for ({top}) skimming flow, ({middle}) wake interference and ({bottom}) isolated roughness. Squares represent $\varepsilon_\text{SVD}=\varepsilon_\text{DMD}={10}^{-3}$ and triangles, ${10}^{-4}$. Red, blue and black correspond to $d=10,20,30$ for SF ($K=225$ snapshots) and WI ($K=215$ snapshots) and $d=5,10$ for IR ($K=90$ snapshots).}
    \label{fig: DMD spectrum callibration}
\end{figure}

The large number of phenomena associated with complex turbulent flows motivates the use of highly-efficient methods to properly identify the behavior of such dynamical structures. This Appendix aims at providing a brief summary of the calibration process concerning the HODMD algorithm. A well-established criteria must be used in order to identify the most robust modes that best characterize the system from the very large number of modes calculated with each variation. Fig.~\ref{fig: DMD spectrum callibration} shows the frequency versus amplitude of the different modes computed using HODMD with a variety of parameters for the three flow regimes. The largest-amplitude and lowest-frequency modes are selected since they possess the more relevant information about the system and the classification will be made based on this distinction. As highlighted in Fig.~\ref{fig: DMD spectrum callibration}, these modes are known to form clusters throughout the spectrum. The amplitude and frequency of the selected modes will be, therefore, the average value of the collection of modes. The number of preserved modes is also a function of the tolerances employed, which were $\varepsilon={10}^{-3}$ and ${10}^{-4}$ in this study. Similarly, the other user-controlled parameter, $d$, changes depending on the number of snapshots to be analyzed. The skimming-flow and wake-interference regimes, with $K=225$ and $215$ snapshots, respectively, were studied with $d=10,20$ and $30$, whilst the $90$ snapshots of the isolated roughness case were studied with $d=5,10$ and $15$. Note that, as mentioned during the theoretical derivation of the Koopman operator (see \S\,\ref{sec: HODMD methodology}), when the number of snapshots is reduced, the value of $d$, which represents the characteristic sliding window process, must also decrease in the same proportion~\cite{LeClainche2020}.

Furthermore, the relative error obtained in the calculations remains fenced in the set of tolerances used. The goal of this study, however, is to identify the largest-amplitude modes in order to provide a broad description of the fundamental patterns driving the flow, rather than to build any accurate reduced-order models based on the physical knowledge of the flow. As a result, the relative error has not be examined further in this study. The reader is referred to {Vega and Le Clainche~\cite{LeClainche2020}} for a more detailed explanation of the calibration process and the influence of the error in the solution.

It is worth noting that when the distance between the obstacles increases, the flow complexity decreases, allowing a smaller number of snapshots to be used to estimate the same flow behavior. Note as well that, despite the fact that the number of snapshots in the isolated-roughness regime is smaller, the time span covered is of the same order of magnitude, as the time step between snapshots has been increased. This is particularly important when dealing with computationally expensive data, but one should bear in mind that in order to accurately represent smaller turbulent scales, a larger number of snapshots with a shorter time step should be utilized instead. However, in general, the HODMD provides a fair balance of computational cost and accuracy.

\section*{Acknowledgments}

RV acknowledges the financial support of the G\"oran Gustafsson foundation. The computations carried out in this study were made possible by resources provided by the Swedish National Infrastructure for Computing (SNIC). AMS and SH were funded by Contract
No. PID2021-128676OB-I00 of Ministerio de Ciencia,
innovaci\'on y Universidades/FEDER. AC and SLC acknowledge the grant PID2020-114173RB-I00 funded by MCIN/AEI/10.13039/501100011033.

\section*{CRediT authorship contribution statement}

\textbf{Álvaro Martínez-Sánchez}: Data curation, Formal analysis, Investigation, Validation, Writing -- original draft, Writing -- review \& editing, Visualization. \textbf{Eneko Lazpita}: Formal analysis, Investigation, Validation, Writing -- original draft, Visualization. \textbf{Adrián Corrochano}: Formal analysis, Investigation, Validation. \textbf{Soledad Le Clainche}: Conceptualization, Funding acquisition, Investigation, Methodology, Software, Writing -- review \& editing. \textbf{Sergio Hoyas}: Conceptualization, Funding acquisition, Investigation, Software, Writing -- review \& editing. \textbf{Ricardo Vinuesa}: Conceptualization, Funding acquisition, Investigation, Methodology, Project administration, Supervision, Writing -- review \& editing. 

\bibliographystyle{elsarticle-num-names}
\bibliography{main}

\begin{thebibliography}{57}
\expandafter\ifx\csname natexlab\endcsname\relax\def\natexlab#1{#1}\fi
\providecommand{\url}[1]{\texttt{#1}}
\providecommand{\href}[2]{#2}
\providecommand{\path}[1]{#1}
\providecommand{\DOIprefix}{doi:}
\providecommand{\ArXivprefix}{arXiv:}
\providecommand{\URLprefix}{URL: }
\providecommand{\Pubmedprefix}{pmid:}
\providecommand{\doi}[1]{\href{http://dx.doi.org/#1}{\path{#1}}}
\providecommand{\Pubmed}[1]{\href{pmid:#1}{\path{#1}}}
\providecommand{\bibinfo}[2]{#2}
\ifx\xfnm\relax \def\xfnm[#1]{\unskip,\space#1}\fi
\bibitem[{Hunt et~al.(1978)Hunt, Abell, Peterka, and Woo}]{Hunt1978}
\bibinfo{author}{J.~C.~R. Hunt}, \bibinfo{author}{C.~J. Abell},
  \bibinfo{author}{J.~A. Peterka}, \bibinfo{author}{H.~Woo},
\newblock \bibinfo{title}{Kinematical studies of the flows around free or
  surface-mounted obstacles; applying topology to flow visualization},
\newblock \bibinfo{journal}{Journal of Fluid Mechanics} \bibinfo{volume}{86}
  (\bibinfo{year}{1978}) \bibinfo{pages}{179–200}.
  \DOIprefix\doi{10.1017/S0022112078001068}.
\bibitem[{Oke(1988)}]{Oke1988}
\bibinfo{author}{T.~Oke},
\newblock \bibinfo{title}{Street design and urban canopy layer climate},
\newblock \bibinfo{journal}{Energy and Buildings} \bibinfo{volume}{11}
  (\bibinfo{year}{1988}) \bibinfo{pages}{103--113}.
  \DOIprefix\doi{10.1016/0378-7788(88)90026-6}.
\bibitem[{Zajic et~al.(2011)Zajic, Fernando, Calhoun, Princevac, Brown, and
  Pardyjak}]{Zajic2011}
\bibinfo{author}{D.~Zajic}, \bibinfo{author}{H.~J.~S. Fernando},
  \bibinfo{author}{R.~Calhoun}, \bibinfo{author}{M.~Princevac},
  \bibinfo{author}{M.~J. Brown}, \bibinfo{author}{E.~R. Pardyjak},
\newblock \bibinfo{title}{Flow and turbulence in an urban canyon},
\newblock \bibinfo{journal}{Journal of Applied Meteorology and Climatology}
  \bibinfo{volume}{50} (\bibinfo{year}{2011}) \bibinfo{pages}{203 -- 223}.
  \DOIprefix\doi{10.1175/2010JAMC2525.1}.
\bibitem[{Heaviside et~al.(2016)Heaviside, Vardoulakis, and
  Cai}]{Heaviside2016}
\bibinfo{author}{C.~Heaviside}, \bibinfo{author}{S.~Vardoulakis},
  \bibinfo{author}{X.-M. Cai},
\newblock \bibinfo{title}{Attribution of mortality to the urban heat island
  during heatwaves in the west midlands, uk},
\newblock \bibinfo{journal}{Environmental Health} \bibinfo{volume}{15}
  (\bibinfo{year}{2016}) \bibinfo{pages}{S27}.
  \DOIprefix\doi{10.1186/s12940-016-0100-9}.
\bibitem[{{European Environment Agency}(2019)}]{EUUrbWorld}
\bibinfo{author}{{European Environment Agency}}, \bibinfo{title}{Air quality in
  Europe -- 2019 report}, \bibinfo{type}{Technical Report}, European Union,
  \bibinfo{year}{2019}. \DOIprefix\doi{10.2800/822355}.
\bibitem[{Becker et~al.(2002)Becker, Lienhart, and Durst}]{Becker2002}
\bibinfo{author}{S.~Becker}, \bibinfo{author}{H.~Lienhart},
  \bibinfo{author}{F.~Durst},
\newblock \bibinfo{title}{Flow around three-dimensional obstacles in boundary
  layers},
\newblock \bibinfo{journal}{Journal of Wind Engineering and Industrial
  Aerodynamics} \bibinfo{volume}{90} (\bibinfo{year}{2002})
  \bibinfo{pages}{265--279}. \DOIprefix\doi{10.1016/S0167-6105(01)00209-4},
  \bibinfo{note}{bluff Body Aerodynamics and Applications}.
\bibitem[{Zhu et~al.(2017)Zhu, Wang, Wang, and Wang}]{zhu2017}
\bibinfo{author}{H.-Y. Zhu}, \bibinfo{author}{C.-Y. Wang},
  \bibinfo{author}{H.-P. Wang}, \bibinfo{author}{J.-J. Wang},
\newblock \bibinfo{title}{Tomographic {PIV} investigation on {3D} wake
  structures for flow over a wall-mounted short cylinder},
\newblock \bibinfo{journal}{Journal of Fluid Mechanics} \bibinfo{volume}{831}
  (\bibinfo{year}{2017}) \bibinfo{pages}{743–778}.
  \DOIprefix\doi{10.1017/jfm.2017.647}.
\bibitem[{Bourgeois et~al.(2012)Bourgeois, Sattari, and
  Martinuzzi}]{Bourgeois2012}
\bibinfo{author}{J.~A. Bourgeois}, \bibinfo{author}{P.~Sattari},
  \bibinfo{author}{R.~J. Martinuzzi},
\newblock \bibinfo{title}{{Coherent vortical and straining structures in the
  finite wall-mounted square cylinder wake}},
\newblock \bibinfo{journal}{International Journal of Heat and Fluid Flow}
  \bibinfo{volume}{35} (\bibinfo{year}{2012}) \bibinfo{pages}{130--140}.
  \DOIprefix\doi{10.1016/j.ijheatfluidflow.2012.01.009}.
\bibitem[{Oertel(1990)}]{Oertel1990}
\bibinfo{author}{H.~Oertel},
\newblock \bibinfo{title}{Wakes behind blunt bodies},
\newblock \bibinfo{journal}{Annual Review of Fluid Mechanics}
  \bibinfo{volume}{22} (\bibinfo{year}{1990}) \bibinfo{pages}{539--562}.
  \DOIprefix\doi{10.1146/annurev.fl.22.010190.002543}.
\bibitem[{Zdravkovich(1997)}]{zdravkovich1997}
\bibinfo{author}{M.~Zdravkovich}, \bibinfo{title}{Flow Around Circular
  Cylinders: Volume 2: Applications}, Flow Around Circular Cylinders: A
  Comprehensive Guide Through Flow Phenomena, Experiments, Applications,
  Mathematical Models, and Computer Simulations, \bibinfo{publisher}{OUP
  Oxford}, \bibinfo{year}{1997}.
\bibitem[{Luo et~al.(2003)Luo, Chew, and Ng}]{Luo2003}
\bibinfo{author}{S.~C. Luo}, \bibinfo{author}{Y.~T. Chew},
  \bibinfo{author}{Y.~T. Ng},
\newblock \bibinfo{title}{Characteristics of square cylinder wake transition
  flows},
\newblock \bibinfo{journal}{Physics of Fluids} \bibinfo{volume}{15}
  (\bibinfo{year}{2003}) \bibinfo{pages}{2549--2559}.
  \DOIprefix\doi{10.1063/1.1596413}.
\bibitem[{Luo et~al.(2007)Luo, Tong, and Khoo}]{Luo2007}
\bibinfo{author}{S.~Luo}, \bibinfo{author}{X.~Tong}, \bibinfo{author}{B.~Khoo},
\newblock \bibinfo{title}{Transition phenomena in the wake of a square
  cylinder},
\newblock \bibinfo{journal}{Journal of Fluids and Structures}
  \bibinfo{volume}{23} (\bibinfo{year}{2007}) \bibinfo{pages}{227--248}.
  \DOIprefix\doi{https://doi.org/10.1016/j.jfluidstructs.2006.08.012}.
\bibitem[{Wang and Zhou(2009)}]{wang_zhou_2009}
\bibinfo{author}{H.~F. Wang}, \bibinfo{author}{Y.~Zhou},
\newblock \bibinfo{title}{The finite-length square cylinder near wake},
\newblock \bibinfo{journal}{Journal of Fluid Mechanics} \bibinfo{volume}{638}
  (\bibinfo{year}{2009}) \bibinfo{pages}{453–490}.
  \DOIprefix\doi{10.1017/S0022112009990693}.
\bibitem[{Bourgeois et~al.(2011)Bourgeois, Sattari, and
  Martinuzzi}]{Bourgeois2011}
\bibinfo{author}{J.~A. Bourgeois}, \bibinfo{author}{P.~Sattari},
  \bibinfo{author}{R.~J. Martinuzzi},
\newblock \bibinfo{title}{Alternating half-loop shedding in the turbulent wake
  of a finite surface-mounted square cylinder with a thin boundary layer},
\newblock \bibinfo{journal}{Physics of Fluids} \bibinfo{volume}{23}
  (\bibinfo{year}{2011}) \bibinfo{pages}{095101}.
  \DOIprefix\doi{10.1063/1.3623463}.
\bibitem[{Monnier et~al.(2018)Monnier, Goudarzi, Vinuesa, and
  Wark}]{Monnier2018}
\bibinfo{author}{B.~Monnier}, \bibinfo{author}{S.~A. Goudarzi},
  \bibinfo{author}{R.~Vinuesa}, \bibinfo{author}{C.~Wark},
\newblock \bibinfo{title}{Turbulent structure of a simplified urban fluid flow
  studied through stereoscopic particle image velocimetry},
\newblock \bibinfo{journal}{Boundary-Layer Meteorology} \bibinfo{volume}{166}
  (\bibinfo{year}{2018}) \bibinfo{pages}{239--268}.
  \DOIprefix\doi{10.1007/s10546-017-0303-9}.
\bibitem[{Sohankar et~al.(1999)Sohankar, Norberg, and Davidson}]{Sohankar1999}
\bibinfo{author}{A.~Sohankar}, \bibinfo{author}{C.~Norberg},
  \bibinfo{author}{L.~Davidson},
\newblock \bibinfo{title}{Simulation of three-dimensional flow around a square
  cylinder at moderate {R}eynolds numbers},
\newblock \bibinfo{journal}{Physics of Fluids} \bibinfo{volume}{11}
  (\bibinfo{year}{1999}) \bibinfo{pages}{288--306}.
  \DOIprefix\doi{10.1063/1.869879}.
\bibitem[{Saha et~al.(2003)Saha, Biswas, and Muralidhar}]{Saha2003}
\bibinfo{author}{A.~Saha}, \bibinfo{author}{G.~Biswas},
  \bibinfo{author}{K.~Muralidhar},
\newblock \bibinfo{title}{Three-dimensional study of flow past a square
  cylinder at low {R}eynolds numbers},
\newblock \bibinfo{journal}{International Journal of Heat and Fluid Flow}
  \bibinfo{volume}{24} (\bibinfo{year}{2003}) \bibinfo{pages}{54--66}.
  \DOIprefix\doi{https://doi.org/10.1016/S0142-727X(02)00208-4}.
\bibitem[{Vinuesa et~al.(2015)Vinuesa, Schlatter, Malm, Mavriplis, and
  Henningson}]{Vinuesa2015}
\bibinfo{author}{R.~Vinuesa}, \bibinfo{author}{P.~Schlatter},
  \bibinfo{author}{J.~Malm}, \bibinfo{author}{C.~Mavriplis},
  \bibinfo{author}{D.~S. Henningson},
\newblock \bibinfo{title}{Direct numerical simulation of the flow around a
  wall-mounted square cylinder under various inflow conditions},
\newblock \bibinfo{journal}{Journal of Turbulence} \bibinfo{volume}{16}
  (\bibinfo{year}{2015}) \bibinfo{pages}{555--587}.
  \DOIprefix\doi{10.1080/14685248.2014.989232}.
\bibitem[{Meinders(1998)}]{Meinders1998}
\bibinfo{author}{E.~Meinders}, \bibinfo{title}{Experimental study of heat
  transfer in turbulent flows over wall-mounted cubes},
  \bibinfo{type}{Dissertation thesis}, TU Delft,
  \bibinfo{address}{Netherlands}, \bibinfo{year}{1998}.
\bibitem[{Jeong and Hussain(1995)}]{jeong_hussain_1995}
\bibinfo{author}{J.~Jeong}, \bibinfo{author}{F.~Hussain},
\newblock \bibinfo{title}{On the identification of a vortex},
\newblock \bibinfo{journal}{Journal of Fluid Mechanics} \bibinfo{volume}{285}
  (\bibinfo{year}{1995}) \bibinfo{pages}{69–94}.
  \DOIprefix\doi{10.1017/S0022112095000462}.
\bibitem[{Sousa(2002)}]{Sousa2002}
\bibinfo{author}{J.~Sousa},
\newblock \bibinfo{title}{Turbulent flow around a surface-mounted obstacle
  using {2D-3C DPIV}},
\newblock \bibinfo{journal}{Experiments in Fluids} \bibinfo{volume}{33}
  (\bibinfo{year}{2002}) \bibinfo{pages}{854--862}.
  \DOIprefix\doi{10.1007/s00348-002-0497-5}.
\bibitem[{Lumley(1967)}]{Lumley1967}
\bibinfo{author}{J.~L. Lumley},
\newblock \bibinfo{title}{The structure of inhomogeneous turbulent flows},
\newblock \bibinfo{journal}{Atmospheric Turbulence and Radio Wave Propagation}
  (\bibinfo{year}{1967}). \URLprefix
  \url{https://ci.nii.ac.jp/naid/10012381873/en/}.
\bibitem[{Schmid(2010)}]{Schmid2010}
\bibinfo{author}{P.~J. Schmid},
\newblock \bibinfo{title}{Dynamic mode decomposition of numerical and
  experimental data},
\newblock \bibinfo{journal}{Journal of Fluid Mechanics} \bibinfo{volume}{656}
  (\bibinfo{year}{2010}) \bibinfo{pages}{5–28}.
  \DOIprefix\doi{10.1017/S0022112010001217}.
\bibitem[{{Le Clainche} and Vega(2017)}]{LeClainche2017b}
\bibinfo{author}{S.~{Le Clainche}}, \bibinfo{author}{J.~M. Vega},
\newblock \bibinfo{title}{Higher order dynamic mode decomposition},
\newblock \bibinfo{journal}{SIAM Journal on Applied Dynamical Systems}
  \bibinfo{volume}{16} (\bibinfo{year}{2017}) \bibinfo{pages}{882--925}.
  \DOIprefix\doi{10.1137/15M1054924}.
\bibitem[{{Le Clainche} et~al.(2017){Le Clainche}, Sastre, Vega, and
  Angel}]{LeClainche2017}
\bibinfo{author}{S.~{Le Clainche}}, \bibinfo{author}{F.~Sastre},
  \bibinfo{author}{J.~M. Vega}, \bibinfo{author}{V.~Angel},
  \bibinfo{title}{Higher order dynamic mode decomposition applied to
  post-process a limited amount of noisy PIV data}, \bibinfo{year}{2017}.
  \DOIprefix\doi{10.2514/6.2017-3304}.
\bibitem[{Le~Clainche and Vega(2017)}]{LeClainche2017c}
\bibinfo{author}{S.~Le~Clainche}, \bibinfo{author}{J.~M. Vega},
\newblock \bibinfo{title}{Higher order dynamic mode decomposition to identify
  and extrapolate flow patterns},
\newblock \bibinfo{journal}{Physics of Fluids} \bibinfo{volume}{29}
  (\bibinfo{year}{2017}) \bibinfo{pages}{084102}.
  \DOIprefix\doi{10.1063/1.4997206}.
\bibitem[{{Le Clainche} et~al.(2017){Le Clainche}, Vega, and
  Soria}]{LECLAINCHE2017d}
\bibinfo{author}{S.~{Le Clainche}}, \bibinfo{author}{J.~M. Vega},
  \bibinfo{author}{J.~Soria},
\newblock \bibinfo{title}{Higher order dynamic mode decomposition of noisy
  experimental data: The flow structure of a zero-net-mass-flux jet},
\newblock \bibinfo{journal}{Experimental Thermal and Fluid Science}
  \bibinfo{volume}{88} (\bibinfo{year}{2017}) \bibinfo{pages}{336--353}.
  \DOIprefix\doi{https://doi.org/10.1016/j.expthermflusci.2017.06.011}.
\bibitem[{{Le Clainche} et~al.(2018){Le Clainche}, P{\'{e}}rez, and
  Vega}]{LeClainche2018b}
\bibinfo{author}{S.~{Le Clainche}}, \bibinfo{author}{J.~M. P{\'{e}}rez},
  \bibinfo{author}{J.~M. Vega},
\newblock \bibinfo{title}{Spatio-temporal flow structures in the
  three-dimensional wake of a circular cylinder},
\newblock \bibinfo{journal}{{IOP} Publishing} \bibinfo{volume}{50}
  (\bibinfo{year}{2018}) \bibinfo{pages}{051406}.
  \DOIprefix\doi{10.1088/1873-7005/aab2f1}.
\bibitem[{Amor et~al.(2020)Amor, P{\'e}rez, Schlatter, Vinuesa, and
  Le~Clainche}]{Amor2020}
\bibinfo{author}{C.~Amor}, \bibinfo{author}{J.~M. P{\'e}rez},
  \bibinfo{author}{P.~Schlatter}, \bibinfo{author}{R.~Vinuesa},
  \bibinfo{author}{S.~Le~Clainche},
\newblock \bibinfo{title}{Soft computing techniques to analyze the turbulent
  wake of a wall-mounted square cylinder},
\newblock \bibinfo{journal}{Advances in Intelligent Systems and Computing}
  \bibinfo{volume}{950} (\bibinfo{year}{2020}) \bibinfo{pages}{577--586}.
  \DOIprefix\doi{10.1007/978-3-030-20055-8_55}.
\bibitem[{Rowley(2005)}]{bpod2005}
\bibinfo{author}{C.~W. Rowley},
\newblock \bibinfo{title}{Model reduction for fluids, using balanced proper
  orthogonal decomposition},
\newblock \bibinfo{journal}{International Journal of Bifurcation and Chaos}
  \bibinfo{volume}{15} (\bibinfo{year}{2005}) \bibinfo{pages}{997--1013}.
  \DOIprefix\doi{10.1142/S0218127405012429}.
\bibitem[{Towne et~al.(2018)Towne, Schmidt, and Colonius}]{spod2018}
\bibinfo{author}{A.~Towne}, \bibinfo{author}{O.~T. Schmidt},
  \bibinfo{author}{T.~Colonius},
\newblock \bibinfo{title}{Spectral proper orthogonal decomposition and its
  relationship to dynamic mode decomposition and resolvent analysis},
\newblock \bibinfo{journal}{Journal of Fluid Mechanics} \bibinfo{volume}{847}
  (\bibinfo{year}{2018}) \bibinfo{pages}{821–867}.
  \DOIprefix\doi{10.1017/jfm.2018.283}.
\bibitem[{{Le Clainche} and Vega(2018)}]{Clainche2018}
\bibinfo{author}{S.~{Le Clainche}}, \bibinfo{author}{J.~M. Vega},
\newblock \bibinfo{title}{Spatio-temporal koopman decomposition},
\newblock \bibinfo{journal}{Journal of Nonlinear Science} \bibinfo{volume}{28}
  (\bibinfo{year}{2018}) \bibinfo{pages}{1793--1842}.
  \DOIprefix\doi{10.1007/s00332-018-9464-z}.
\bibitem[{Le~Clainche et~al.(2020)Le~Clainche, Izbassarov, Rosti, Brandt, and
  Tammisola}]{LeClainche2020}
\bibinfo{author}{S.~Le~Clainche}, \bibinfo{author}{D.~Izbassarov},
  \bibinfo{author}{M.~Rosti}, \bibinfo{author}{L.~Brandt},
  \bibinfo{author}{O.~Tammisola},
\newblock \bibinfo{title}{Coherent structures in the turbulent channel flow of
  an elastoviscoplastic fluid},
\newblock \bibinfo{journal}{Journal of Fluid Mechanics} \bibinfo{volume}{888}
  (\bibinfo{year}{2020}) \bibinfo{pages}{A5}.
  \DOIprefix\doi{10.1017/jfm.2020.31}.
\bibitem[{Le~Clainche et~al.(2022)Le~Clainche, Rosti, and
  Brandt}]{clainche2022}
\bibinfo{author}{S.~Le~Clainche}, \bibinfo{author}{M.~Rosti},
  \bibinfo{author}{L.~Brandt},
\newblock \bibinfo{title}{A data-driven model based on modal decomposition:
  application to the turbulent channel flow over an anisotropic porous wall},
\newblock \bibinfo{journal}{Journal of Fluid Mechanics} \bibinfo{volume}{939}
  (\bibinfo{year}{2022}).
\bibitem[{Mendez et~al.(2021)Mendez, {Le Clainche}, Moreno-Ramos, and
  Vega}]{Mendez2021}
\bibinfo{author}{C.~Mendez}, \bibinfo{author}{S.~{Le Clainche}},
  \bibinfo{author}{R.~Moreno-Ramos}, \bibinfo{author}{J.~M. Vega},
\newblock \bibinfo{title}{A new automatic, very efficient method for the
  analysis of flight flutter testing data},
\newblock \bibinfo{journal}{Aerospace Science and Technology}
  \bibinfo{volume}{114} (\bibinfo{year}{2021}) \bibinfo{pages}{106749}.
  \DOIprefix\doi{https://doi.org/10.1016/j.ast.2021.106749}.
\bibitem[{Lazpita et~al.(2022)Lazpita, Mart{\'\i}nez-S{\'a}nchez, Corrochano,
  Hoyas, Le~Clainche, and Vinuesa}]{laz22}
\bibinfo{author}{E.~Lazpita},
  \bibinfo{author}{{\'A}.~Mart{\'\i}nez-S{\'a}nchez},
  \bibinfo{author}{A.~Corrochano}, \bibinfo{author}{S.~Hoyas},
  \bibinfo{author}{S.~Le~Clainche}, \bibinfo{author}{R.~Vinuesa},
\newblock \bibinfo{title}{On the generation and destruction mechanisms of arch
  vortices in urban fluid flows},
\newblock \bibinfo{journal}{Physics of Fluids} \bibinfo{volume}{34}
  (\bibinfo{year}{2022}) \bibinfo{pages}{051702}.
\bibitem[{Fischer et~al.(2008)Fischer, Lottes, and Kerkemeier}]{Nek5000}
\bibinfo{author}{P.~Fischer}, \bibinfo{author}{J.~Lottes},
  \bibinfo{author}{S.~Kerkemeier}, \bibinfo{title}{{Nek5000}: open source
  spectral element {CFD} solver}, \bibinfo{year}{2008}. \URLprefix
  \url{http://nek5000.mcs.anl.gov}.
\bibitem[{Patera(1984)}]{PATERA1984}
\bibinfo{author}{A.~T. Patera},
\newblock \bibinfo{title}{A spectral element method for fluid dynamics: Laminar
  flow in a channel expansion},
\newblock \bibinfo{journal}{Journal of Computational Physics}
  \bibinfo{volume}{54} (\bibinfo{year}{1984}) \bibinfo{pages}{468--488}.
  \DOIprefix\doi{https://doi.org/10.1016/0021-9991(84)90128-1}.
\bibitem[{Hoyas and Jiménez(2006)}]{Hoyas2006}
\bibinfo{author}{S.~Hoyas}, \bibinfo{author}{J.~Jiménez},
\newblock \bibinfo{title}{Scaling of the velocity fluctuations in turbulent
  channels up to ${Re}_\tau=2003$},
\newblock \bibinfo{journal}{Physics of Fluids} \bibinfo{volume}{18}
  (\bibinfo{year}{2006}) \bibinfo{pages}{011702}.
  \DOIprefix\doi{10.1063/1.2162185}.
\bibitem[{Simens et~al.(2009)Simens, Jiménez, Hoyas, and Mizuno}]{SIMENS2009}
\bibinfo{author}{M.~P. Simens}, \bibinfo{author}{J.~Jiménez},
  \bibinfo{author}{S.~Hoyas}, \bibinfo{author}{Y.~Mizuno},
\newblock \bibinfo{title}{A high-resolution code for turbulent boundary
  layers},
\newblock \bibinfo{journal}{Journal of Computational Physics}
  \bibinfo{volume}{228} (\bibinfo{year}{2009}) \bibinfo{pages}{4218--4231}.
  \DOIprefix\doi{https://doi.org/10.1016/j.jcp.2009.02.031}.
\bibitem[{Negi et~al.(2018)Negi, Vinuesa, Hanifi, Schlatter, and
  Henningson}]{NEGI2018}
\bibinfo{author}{P.~Negi}, \bibinfo{author}{R.~Vinuesa},
  \bibinfo{author}{A.~Hanifi}, \bibinfo{author}{P.~Schlatter},
  \bibinfo{author}{D.~Henningson},
\newblock \bibinfo{title}{Unsteady aerodynamic effects in small-amplitude pitch
  oscillations of an airfoil},
\newblock \bibinfo{journal}{International Journal of Heat and Fluid Flow}
  \bibinfo{volume}{71} (\bibinfo{year}{2018}) \bibinfo{pages}{378--391}.
  \DOIprefix\doi{https://doi.org/10.1016/j.ijheatfluidflow.2018.04.009}.
\bibitem[{Noorani et~al.(2016)Noorani, Vinuesa, Brandt, and
  Schlatter}]{Noorani2016}
\bibinfo{author}{A.~Noorani}, \bibinfo{author}{R.~Vinuesa},
  \bibinfo{author}{L.~Brandt}, \bibinfo{author}{P.~Schlatter},
\newblock \bibinfo{title}{Aspect ratio effect on particle transport in
  turbulent duct flows},
\newblock \bibinfo{journal}{Physics of Fluids} \bibinfo{volume}{28}
  (\bibinfo{year}{2016}) \bibinfo{pages}{115103}.
  \DOIprefix\doi{10.1063/1.4966026}.
\bibitem[{Vinuesa(2021)}]{VINUESA2021}
\bibinfo{author}{R.~Vinuesa},
\newblock \bibinfo{title}{High-fidelity simulations in complex geometries:
  Towards better flow understanding and development of turbulence models},
\newblock \bibinfo{journal}{Results in Engineering} \bibinfo{volume}{11}
  (\bibinfo{year}{2021}) \bibinfo{pages}{100254}.
  \DOIprefix\doi{https://doi.org/10.1016/j.rineng.2021.100254}.
\bibitem[{Tanarro et~al.(2020)Tanarro, Vinuesa, and
  Schlatter}]{tanarro_vinuesa_schlatter_2020}
\bibinfo{author}{A.~Tanarro}, \bibinfo{author}{R.~Vinuesa},
  \bibinfo{author}{P.~Schlatter},
\newblock \bibinfo{title}{Effect of adverse pressure gradients on turbulent
  wing boundary layers},
\newblock \bibinfo{journal}{Journal of Fluid Mechanics} \bibinfo{volume}{883}
  (\bibinfo{year}{2020}) \bibinfo{pages}{A8}.
  \DOIprefix\doi{10.1017/jfm.2019.838}.
\bibitem[{Atzori et~al.(2022)Atzori, Torres, Vidal, {Le Clainche}, Hoyas, and
  Vinuesa}]{marco2022}
\bibinfo{author}{M.~Atzori}, \bibinfo{author}{P.~Torres},
  \bibinfo{author}{A.~Vidal}, \bibinfo{author}{S.~{Le Clainche}},
  \bibinfo{author}{S.~Hoyas}, \bibinfo{author}{R.~Vinuesa},
  \bibinfo{title}{High-resolution large-eddy simulations of simplified urban
  flows}, \bibinfo{year}{2022}. \DOIprefix\doi{10.48550/ARXIV.2207.07210}.
  \href{http://arxiv.org/abs/2207.07210}{{\tt arXiv:2207.07210}}.
\bibitem[{Vinuesa et~al.(2018)Vinuesa, Negi, Atzori, Hanifi, Henningson, and
  Schlatter}]{VINUESA201886}
\bibinfo{author}{R.~Vinuesa}, \bibinfo{author}{P.~Negi},
  \bibinfo{author}{M.~Atzori}, \bibinfo{author}{A.~Hanifi},
  \bibinfo{author}{D.~Henningson}, \bibinfo{author}{P.~Schlatter},
\newblock \bibinfo{title}{Turbulent boundary layers around wing sections up to
  ${Re}_c=1,000,000$},
\newblock \bibinfo{journal}{International Journal of Heat and Fluid Flow}
  \bibinfo{volume}{72} (\bibinfo{year}{2018}) \bibinfo{pages}{86--99}.
  \DOIprefix\doi{https://doi.org/10.1016/j.ijheatfluidflow.2018.04.017}.
\bibitem[{Vinuesa et~al.(2017)Vinuesa, Hosseini, Hanifi, Henningson, and
  Schlatter}]{Vinuesa2017}
\bibinfo{author}{R.~Vinuesa}, \bibinfo{author}{S.~M. Hosseini},
  \bibinfo{author}{A.~Hanifi}, \bibinfo{author}{D.~S. Henningson},
  \bibinfo{author}{P.~Schlatter},
\newblock \bibinfo{title}{Pressure-gradient turbulent boundary layers
  developing around a wing section},
\newblock \bibinfo{journal}{Flow, Turbulence and Combustion}
  \bibinfo{volume}{99} (\bibinfo{year}{2017}) \bibinfo{pages}{613--641}.
  \DOIprefix\doi{10.1007/s10494-017-9840-z}.
\bibitem[{Dong et~al.(2014)Dong, Karniadakis, and Chryssostomidis}]{DONG2014}
\bibinfo{author}{S.~Dong}, \bibinfo{author}{G.~Karniadakis},
  \bibinfo{author}{C.~Chryssostomidis},
\newblock \bibinfo{title}{A robust and accurate outflow boundary condition for
  incompressible flow simulations on severely-truncated unbounded domains},
\newblock \bibinfo{journal}{Journal of Computational Physics}
  \bibinfo{volume}{261} (\bibinfo{year}{2014}) \bibinfo{pages}{83--105}.
  \DOIprefix\doi{https://doi.org/10.1016/j.jcp.2013.12.042}.
\bibitem[{Garratt(1994)}]{garratt1994}
\bibinfo{author}{J.~R. Garratt},
\newblock \bibinfo{title}{The atmospheric boundary layer},
\newblock \bibinfo{journal}{Earth-Science Reviews} \bibinfo{volume}{37}
  (\bibinfo{year}{1994}) \bibinfo{pages}{89--134}.
\bibitem[{Sirovich(1987)}]{Sirovich1987}
\bibinfo{author}{L.~Sirovich},
\newblock \bibinfo{title}{Turbulence and the dynamics of coherent structures
  {P}art {I}: Coherent structures},
\newblock \bibinfo{journal}{Quarterly of Applied Mathematics}
  \bibinfo{volume}{45} (\bibinfo{year}{1987}) \bibinfo{pages}{561--571}.
\bibitem[{Vega and {Le Clainche}(2020)}]{Vega2020Book}
\bibinfo{author}{J.~M. Vega}, \bibinfo{author}{S.~{Le Clainche}},
  \bibinfo{title}{Higher Order Dynamic Mode Decomposition and Its
  Applications}, \bibinfo{publisher}{Elsevier ISBN: 9780128197431},
  \bibinfo{year}{2020}. \DOIprefix\doi{10.1016/c2019-0-00038-6}.
\bibitem[{Le~Clainche et~al.(2022)Le~Clainche, Rosti, and Brandt}]{le2022data}
\bibinfo{author}{S.~Le~Clainche}, \bibinfo{author}{M.~Rosti},
  \bibinfo{author}{L.~Brandt},
\newblock \bibinfo{title}{A data-driven model based on modal decomposition:
  application to the turbulent channel flow over an anisotropic porous wall},
\newblock \bibinfo{journal}{Journal of Fluid Mechanics} \bibinfo{volume}{939}
  (\bibinfo{year}{2022}).
\bibitem[{Zhao et~al.(2021)Zhao, Mamoon, and Wu}]{Zhao2021}
\bibinfo{author}{M.~Zhao}, \bibinfo{author}{A.-A. Mamoon},
  \bibinfo{author}{H.~Wu},
\newblock \bibinfo{title}{Numerical study of the flow past two wall-mounted
  finite-length square cylinders in tandem arrangement},
\newblock \bibinfo{journal}{Physics of Fluids} \bibinfo{volume}{33}
  (\bibinfo{year}{2021}) \bibinfo{pages}{093603}.
  \DOIprefix\doi{10.1063/5.0058394}.
\bibitem[{Meinders and Hanjalić(2002)}]{Meinders2002}
\bibinfo{author}{E.~Meinders}, \bibinfo{author}{K.~Hanjalić},
\newblock \bibinfo{title}{Experimental study of the convective heat transfer
  from in-line and staggered configurations of two wall-mounted cubes},
\newblock \bibinfo{journal}{International Journal of Heat and Mass Transfer}
  \bibinfo{volume}{45} (\bibinfo{year}{2002}) \bibinfo{pages}{465--482}.
  \DOIprefix\doi{https://doi.org/10.1016/S0017-9310(01)00180-6}.
\bibitem[{Pastor et~al.(2012)Pastor, Binda, and Harčarik}]{Pastor2012}
\bibinfo{author}{M.~Pastor}, \bibinfo{author}{M.~Binda},
  \bibinfo{author}{T.~Harčarik},
\newblock \bibinfo{title}{Modal assurance criterion},
\newblock \bibinfo{journal}{Procedia Engineering} \bibinfo{volume}{48}
  (\bibinfo{year}{2012}) \bibinfo{pages}{543--548}.
  \DOIprefix\doi{https://doi.org/10.1016/j.proeng.2012.09.551},
  \bibinfo{note}{modelling of Mechanical and Mechatronics Systems}.
\bibitem[{Torres et~al.(2021)Torres, Le~Clainche, and Vinuesa}]{energies2021}
\bibinfo{author}{P.~Torres}, \bibinfo{author}{S.~Le~Clainche},
  \bibinfo{author}{R.~Vinuesa},
\newblock \bibinfo{title}{On the experimental, numerical and data-driven
  methods to study urban flows},
\newblock \bibinfo{journal}{Energies} \bibinfo{volume}{14}
  (\bibinfo{year}{2021}). \DOIprefix\doi{10.3390/en14051310}.
\bibitem[{Cooley and Tukey(1965)}]{cooley1965}
\bibinfo{author}{J.~W. Cooley}, \bibinfo{author}{J.~W. Tukey},
\newblock \bibinfo{title}{An algorithm for the machine calculation of complex
  {F}ourier series},
\newblock \bibinfo{journal}{Mathematics of computation} \bibinfo{volume}{19}
  (\bibinfo{year}{1965}) \bibinfo{pages}{297--301}.

\end{thebibliography}

\end{document}